\def\kms{\hbox{km s$^{-1}$~}}

\def\lesssim{\mathrel{\hbox{\rlap{\hbox{\lower2pt\hbox{$\sim$}}}\raise2pt\hbox{$<$}}}}

\documentclass[useAMS,usenatbib]{mn2e}

\title[The NLR of the nearest obscured quasar]{Deconstructing the narrow-line region of the nearest obscured quasar}
\author[Villar Mart\'\i n et al.]{M. Villar Mart\'\i n$^{1,2}$, E. Bellocchi$^{1}$, J. Stern$^3$, C. Ramos Almeida$^{4,5}$, C. Tadhunter$^6$\\
\newauthor  R. Gonz\' alez Delgado$^7$ \\
$^1$Centro de Astrobiolog\'\i a (INTA-CSIC), Carretera de Ajalvir, km 4, E-28850 Torrej\'on de Ardoz, Madrid, Spain.  villarmm@cab.inta-csic.es \\
$^2$Astro-UAM, UAM, Unidad Asociada CSIC, Facultad de Ciencias, Campus de Cantoblanco, E-28049, Madrid, Spain \\
$^3$Max  Planck Institute f\"ur Astronomie, K\"onigstuhl 17, D-69117 Heidelberg, Germany \\
$^4$Instituto de Astrof\'\i sica de Canarias, C/ V\'\i a L\'actea, s/n, E38205  La Laguna,Tenerife, Spain \\
$^5$Departamento de Astrof\'\i sica, Universidad de La Laguna, E-38206, La Laguna, Tenerife, Spain \\
$^6$Department of Physics and Astronomy, University of Sheffield, Sheffield S3 7RH, UK \\
$^7$Instituto de Astrof\'\i sica de Andaluc\'\i a (CSIC), Glorieta de la Astronom\'\i a, s/n, 18008 Granada,  Spain \\
}
\begin{document}

\date{Accepted 2015 August 11. Received 2015 July 14; in original form 2015 January 18}

\pagerange{\pageref{firstpage}--\pageref{lastpage}} \pubyear{2002}

\maketitle

\label{firstpage}

\begin{abstract}
 We study the physical and kinematic properties of the  narrow line region (NLR) of the nearest obscured quasar MRK~477  ($z$=0.037), using optical and near-infrared (NIR) spectroscopy. 
About 100  emission lines  are identified in the optical+NIR  spectrum (90 in the optical), including several narrow {\it optical}  Fe$^+$  lines. To our knowledge, this is the first type 2 active galactic nucleus (AGN) with such a detection.  The Fe$^+$ lines can be explained as the natural emission from the NLR photoionized by the AGN.  Coronal line emission can only be confirmed in the NIR spectrum.

As in many other AGN, a significant correlation is found between the lines full width at half-maximum and the critical density log($n_{\rm crit}$).
We propose that it is caused by the outflow. This could be the case in other AGNs.

 The  nuclear jet-induced ionized outflow  has been kinematically isolated in many emission lines  covering a broad range of ionization potentials and critical densities.  
 It  is concentrated    within  $R\sim$few$\times$100 pc from the central engine. 
The outflowing gas is  denser ($n\ga$8 000 cm$^{-3}$) than the ambient non-perturbed gas ($n\sim$400-630 cm$^{-3}$). This could be  due to the compression effect of the jet induced shocks. 
Alternatively, we propose that the outflow has been triggered by the jet  at $R\la$220 pc  (possibly at $\la$30 pc)   and we trace how the impact  weakens as it propagates outwards 
  following the radiation-pressure dominated density gradient.

The different  kinematic behaviour of [FeII]$\lambda$1.644 $\mu$m suggests that its emission is enhanced by shocks induced by the nuclear outflow/jet and is preferentially emitted  at a different, less reddened spatial location.

\end{abstract}


\section{Introduction}

MRK~477  (SDSS J144038.1+533016, $z$=0.037 and luminosity distance $D_L$=161 Mpc) is a type 2 luminous active galactic nucleus (AGN). Also known as  I Zw 92, it was first identified by 
 \cite{zwi66} as a compact galaxy. It is usually referred to as the most luminous Seyfert 2 in the local Universe. As pointed out by   \cite{hec97}, 
it has the highest [O III]$\lambda$5007 luminosity ($L_{\rm [OIII]}$=3.3$\times$10$^{42}$ erg  s$^{-1}$)  of any of the 140 Seyfert nuclei (type 1 or 2) compiled
by  \cite{whi92} and the fifth highest radio power ($P_{\rm 1.4 GHz}=$2.0$\times$10$^{30}$ erg s$^{-1}$ Hz$^{-1}$ at 1.4 GHz). Indeed, 
its high $L_{\rm [OIII]}$ places it in the regime of optically selected obscured quasars
(QSO2), according to the selection criteria  defined by   \cite{zak03}, $L_{\rm [OIII]}>$1.2$\times$10$^{42}$ erg  s$^{-1}$.  The   [OIII] luminosity   implies a bolometric luminosity $L_{bol}$=8.6$\times$10$^{45}$ erg  s$^{-1}$ (Stern \& Laor \citeyear{stern12}; see also Lamastra et al.  \citeyear{lam09}), which is in the quasar range (Shen et al.  \citeyear{shen11}).

\begin{figure*}
\includegraphics{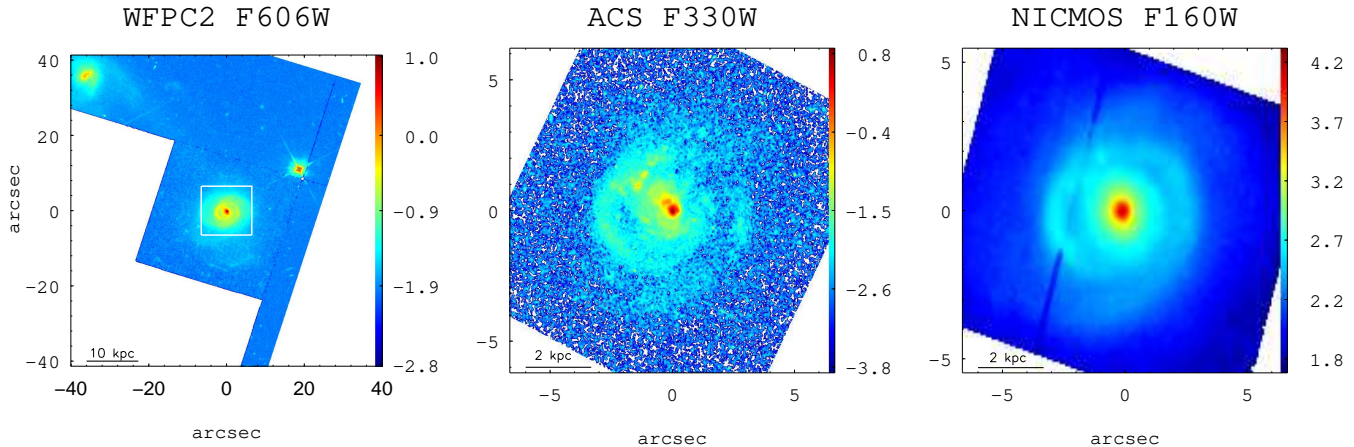}
\vspace{2.6in}
\caption{Hubble Space Telescope (HST) images of MRK~477. They were obtained with Wide-Field Planetary Camera 2 (WFPC2, F606W filter, left), the Advanced Camera for Surveys (ACS/HRC, F330W filter, middle) and the near-infrared Camera and Multi-Object Spectrometer (NICMOS, F160W filter, right) respectively. The flux intensity maps are represented in logarithmic scale such that F [erg s$^{-2}$ cm$^{-2}$ \AA$^{-1}$  arcsec$^{-2}$] = 10$^{A+b}$, where $A$ is a constant factor (i.e., -18.3, -14.5 and -16.4, respectively, for WFPC2, ACS and NICMOS images) and $b$ is indicated in the coloured bar. The white box in the WFPC2 image identifies the FoV covered by the ACS and NICMOS images. The horizontal line at the bottom left of these images corresponds to a scale of 10 kpc for the WFPC2 image and to a scale of 2 kpc for the ACS and NICMOS images. The spatial scale for this galaxy is 0.725 kpc/ arcsec . North is at the top and East to the left in all the panels.}             
\label{fig:hst}
\end{figure*}

 The quasar
 host galaxy is interacting with an emission line companion, possibly a LINER,  located 50 arcsec ($\sim$36 kpc) to the north   (De Robertis et al. \citeyear{derob87}).  Hubble Space Telescope (HST) images
 are shown in Fig. \ref{fig:hst}\footnote{The HST images  were retrieved from the HST Science Legacy Archive. They are part of the  programmes 9379 with principal investigator PI H. Schmitt (ACS/HRC image),  8597 with PI M. Michael Regan (WFPC image) and 7330 with PI J. Mulchaey  (NICMOS image).}. Both galaxies are connected by a faint bridge and prominent tidal tails are apparent.    
 The ratio of stellar masses between the quasar host and the companion is  M1/M2=1.6 (Koss et al. \citeyear{koss12}).

  The quasar host shows a compact blue central source,  peculiar among type 2 AGNs, but no evidence for a broad line region (BLR),  prompting some authors to suggest that the quasar lacks one.     However, polarization observations have revealed a hidden BLR (e.g.  Tran et al. \citeyear{tran92}; Tran \citeyear{tran95}; Shu et al. \citeyear{shu07}). The nuclear continuum  polarization is $\la$1.5\%, significantly lower than that of the broad lines, implying that the majority of the optical continuum cannot be due to scattered light from the hidden type 1 nucleus. 
   The blue central  source is instead  associated with a compact dusty starburst (effective radius$<$0.2 kpc) which occurred  $\sim$6 Myr ago in a brief period of time and whose  light dominates from the ultraviolet  to the near-infrared  (NIR; Heckman et al. \citeyear{hec97}, Gonz\'alez Delgado et al. \citeyear{gon98}). This is therefore a starburst-AGN hybrid system.  MRK~477 has  a star-forming rate SFR=24 M$_{\odot}$ yr$^{-1}$ inferred from the infrared (IR)  luminosity (assuming that  it is starburst-dominated) 
log$\frac{L_{\rm IR}}{L_{\odot}}$=11.14  (Krug et al. \citeyear{krug10}). This places it  in the regime of luminous IR galaxies (LIRGs: 11$\le {\rm log}\frac{L_{\rm IR}}{L_{\odot}} <$12). The detection of the Wolf Rayet (WR) blue bump around the He II$\lambda$4686 line by \cite{hec97} led these authors to propose that   MRK~477 is a luminous member of the class  of WR  galaxies.  

 The QSO2 host  harbors a supermassive black hole with a mass in the range log($M_{\rm  BH}$)$\sim$7.18  (as derived from the  $M_{\rm  BH}$ versus  $\sigma_*$ correlation) to 8.84  (as inferred from the polarimetric H$\beta$ line; Zhang, Bian \& Huang \citeyear{zhang08}).

Although radio quiet according to the $L_{\rm [OIII)}$ versus  $L_{\rm rad}$, radio loud versus  radio quiet, classification criteria (explained in Villar-Mart\'\i n et al.  \citeyear{vm14}),  MRK~477  shows a clear excess of radio emission compared to that expected from the stellar contribution. The
 8.4 GHz Very Large Array (VLA) radio continuum map (0.26 arcsec  resolution)
shows a triple radio source ($\sim$1.2 arcsec  total extension) whose morphology correlates with that of the narrow line region (NLR). The size and overall northeast-southwest axis of the radio structures are shared by the [OIII] emission, which extends  up to a  similar distance  to the north-east (Heckman et al. \citeyear{hec97}). 
 
\cite{shu81} reported for the first time that the emission lines have a large blue-ward asymmetry (i.e., excess)
at low intensity level relative to the continuum. \cite{vm14} proposed that the ionized nuclear outflow responsible for such an asymmetry has been triggered by the interaction between the radio source and the NLR. These authors proposed that negative feedback can be triggered by the radio structures in a significant fraction of radio quiet quasars.  Thanks to its high intrinsic luminosity and closeness,
MRK~477 is an excellent test object to gain further insight into this mechanism.

We present here a detailed study of the optical and near-infrared  spectra of MRK~477. We explore a diversity of aspects  that  provide a more complete understanding of the nature of this object, example of a type 2 quasar in the nearby universe, as well as a starburst-AGN hybrid system.
The paper is organized as follows. The  spectra are described in Sect. 2, together with the spectral  fitting procedure.  We present results in Sect. 3 regarding  line identification, correlations between the gas kinematics and a diversity of parameters (ionization potential, critical density),  the spatial extension of the ionized gas, the ionized outflow, the presence of narrow Fe$^+$ emission lines, the coronal line spectrum and the detection of WR features. The results are discussed in Sect. 4 and the conclusions are presented in Sect. 5.

We assume
$\Omega_{\Lambda}$=0.7, $\Omega_{\rm M}$=0.3, $H_0$=71 km s$^{-1}$ Mpc$^{-1}$.  At $z=$0.037, 1 arcsec  corresponds  0.725 kpc
(Wright \citeyear{wright06}).

\section{Data set}

\subsection{Optical spectrum}

The optical spectrum was obtained as part of the Sloan Digital Sky Survey (SDSS, York et al. {\citeyear{york00}). It   spans the rest frame range $\sim$3660-8880 \AA\ range (Figs. \ref{fig:spectra} and  \ref{fig:spectrab}). 
 It corresponds to an aperture defined by 
the 3 arcsec  diameter SDSS fibre ($\sim$2.2 kpc at $z=$0.037) centred at the galaxy nucleus. For comparison, the triple radio source identified by \cite{hec97} has
a total extension of $\sim$1.2 arcsec  and thus it is well within the fibre area.

The spectral resolution is $\sim$180$\pm$20 km s$^{-1}$.  All values of emission line full width at half-maximum (FWHM) have been corrected for instrumental broadening.  

The main emission lines  have large equivalent widths and the stellar features  are comparatively weak.  Underlying stellar absorption of the Balmer lines is expected to be negligible and therefore subtracting  the stellar continuum is not necessary for our purposes. 

\subsection{NIR spectrum: observations and data reduction}

NIR H+K long-slit spectra    were obtained with the NIR camera/spectrometer LIRIS (Long-slit Intermediate Resolution Infrared Spectrograph;  Acosta-Pulido et al. \citeyear{aco03}; Manchado et al. \citeyear{man04}), attached to the Cassegrain focus of the 4.2 m William
Herschel Telescope (WHT). LIRIS is equipped with a Rockwell Hawaii 1024$\times$1024 HgCdTe array detector, whose spatial scale is
0.25 arcsec  pixel$^{-1}$.  

The NIR observations were performed as part of  the Isaac Newton Group  service programme (programme SW2014b22) in two different nights,    28 March 2015 and 02 May 2015. The H+K grism was used. It covers the  1.388-2.419 $\mu$m spectral range and provides a dispersion of 9.7\AA\ pix$^{-1}$. The seeing size during the  March observations was FWHM$\sim$2\arcsec  (as measured from the standard star). 
A 1\arcsec~ wide slit was used oriented at position angle PA=43$\degr$  East of North.  
This PA is aligned with the radio axis and the NLR axis as seeing in the HST image presented by \cite{hec97}. The resulting spectral resolution was 35.7$\pm$2.2 \AA\ (or 550$\pm$34 km s$^{-1}$ at the observed wavelength of Pa$\alpha$).  In Fig. \ref{fig:liris} we show the NIR spectrum extracted from a 3.5\arcsec ~ aperture, centred at the position of the nucleus. The total exposure time on source was 6400 s (16$\times$400 s). 

The May data were obtained with  seeing FWHM=0.65$\pm$0.05\arcsec (measured from the standard star). A 0.75\arcsec~slit was used, also with PA=43$\degr$. The resulting spectral resolution was 23.2$\pm$1.2 \AA\   (or 318$\pm$18 km s$^{-1}$). The observations were performed through clouds  and  the signal to noise ratio of the final spectrum was low.  In spite of this, the good seeing conditions allowed   
to obtain more accurate information on the spatial extension of the strongest emission line Pa$\alpha$. The total exposure time on source was 4000 s (10$\times$400 s).  All FWHM values of the NIR spectra have been corrected for instrumental broadening.

The data were reduced following standard procedures for near-IR spectroscopy, using the {\sc lirisdr} dedicated software within the {\sc IRAF}\footnote{{\sc IRAF} is distributed by the National Optical Astronomy Observatory, which is operated by the Association of Universities for the Research in Astronomy, Inc., under cooperative agreement with the National Science Foundation (http://iraf.noao.edu/).} enviroment. For a detailed description of the reduction
process see \cite{ram09}. Consecutive pairs of AB two-dimensional spectra were subtracted to remove the sky background. The resulting frames were then wavelength calibrated and flat-fielded before registering and co-adding all frames to provide the final spectra.

The absolute flux  calibration of the March spectrum is intended to be an approximation since the spectra of the comparison star are likely subjected to slit losses due to
centring and tracking errors.  We note that in this work we refer to flux ratios only, and therefore the uncertainty in the absolute flux calibration of the spectra does not affect our results.

\begin{figure*}
\includegraphics{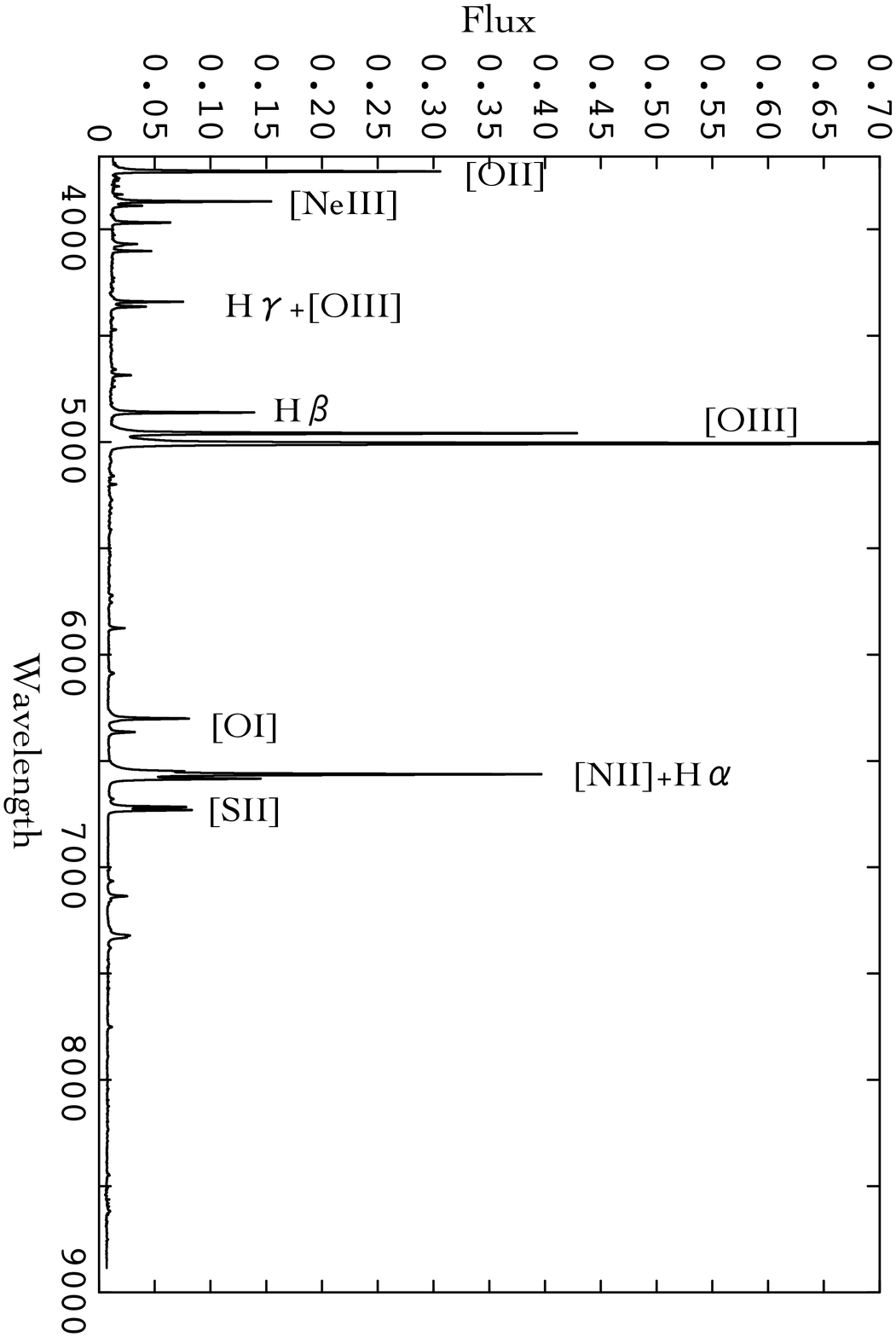}
\vspace{3.3in}
\caption{SDSS spectrum of MRK~477. Flux is given in units of  $\times$10$^{-13}$ erg s$^{-1}$ cm$^{-2}$ \AA$^{-1}$ and the rest-frame wavelength is in \AA. Some of the strongest emission lines are indicated.}
\label{fig:spectra}
\includegraphics{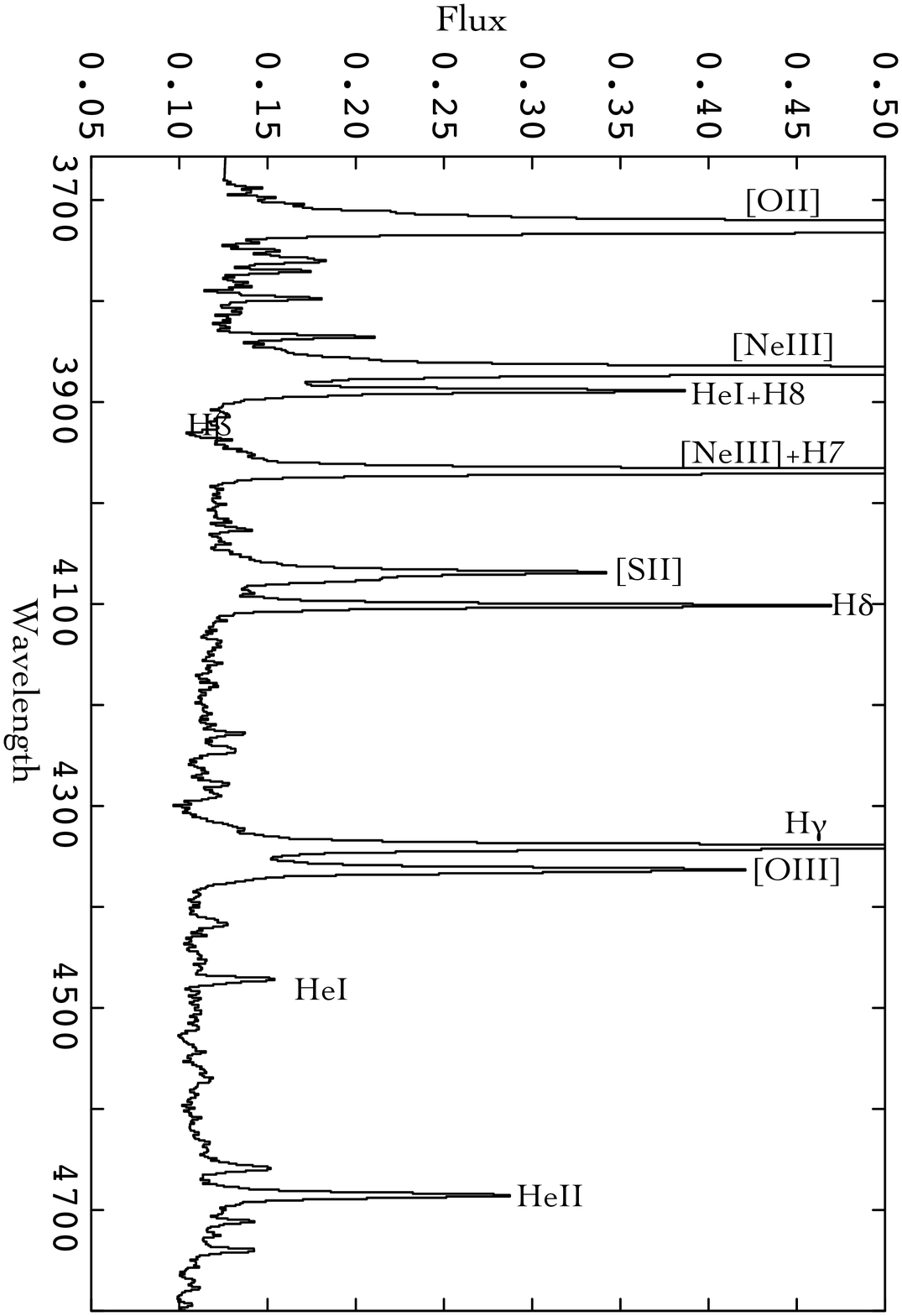}
\includegraphics{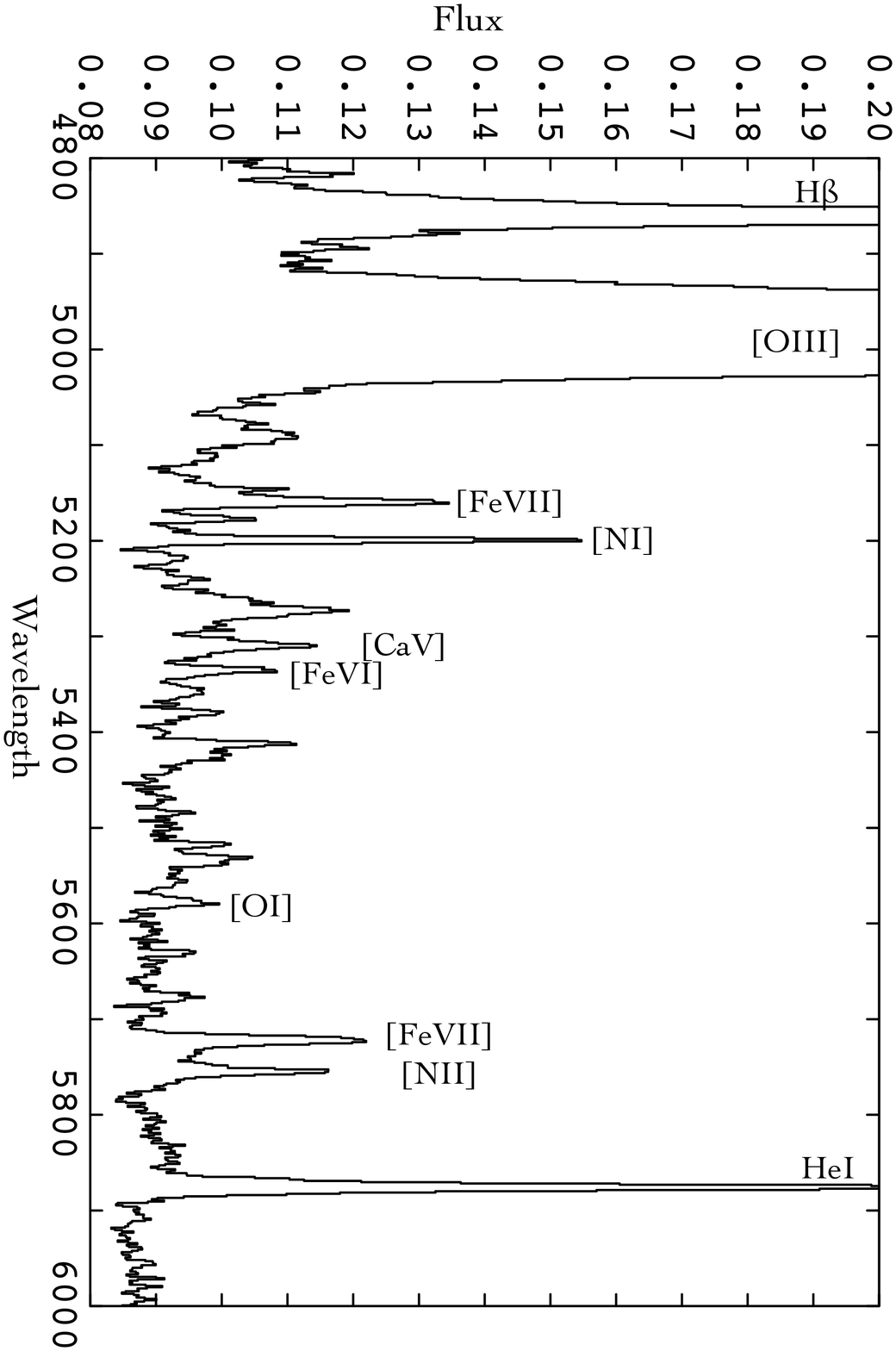}
\vspace{2.7in}
\includegraphics{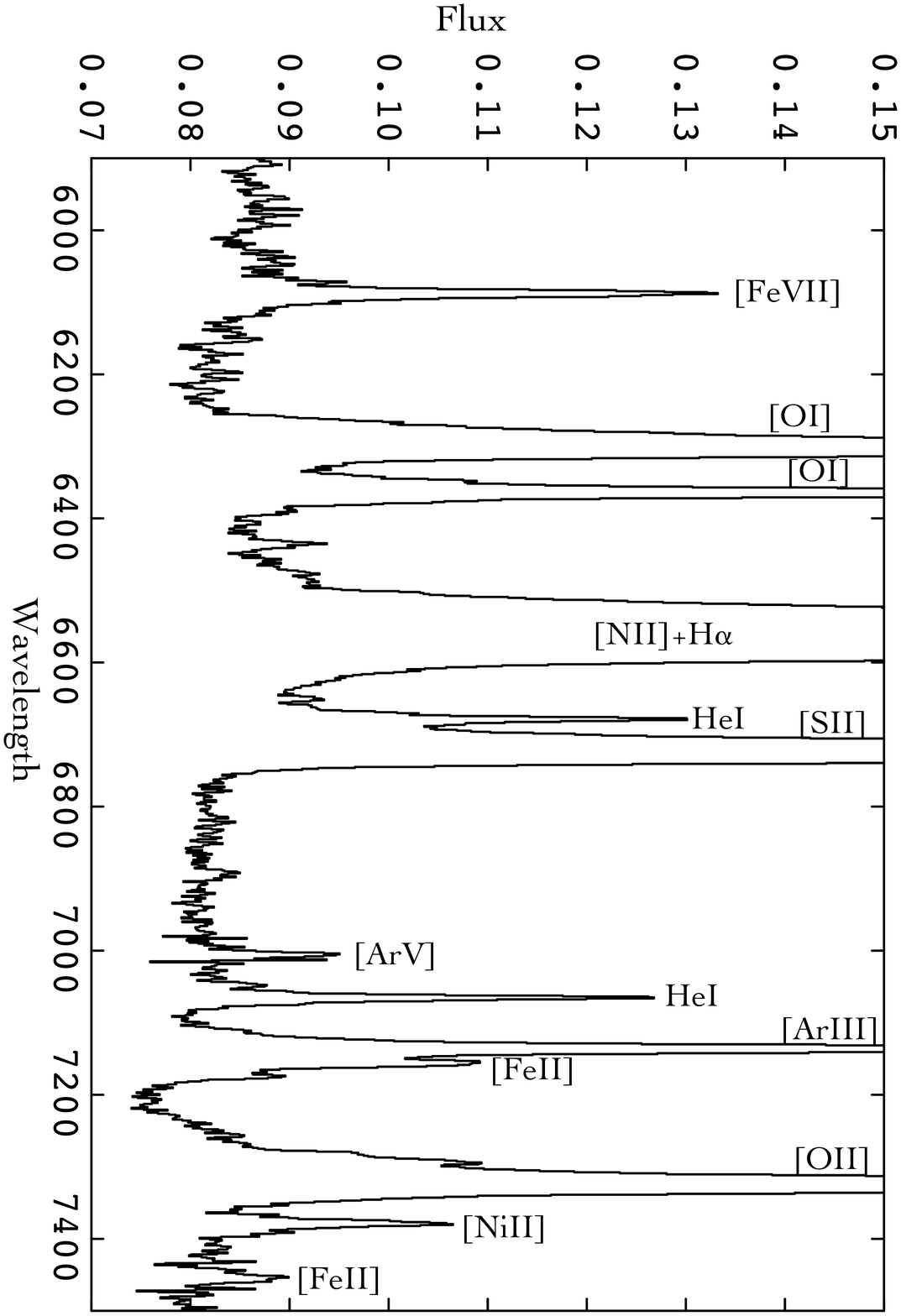}
\includegraphics{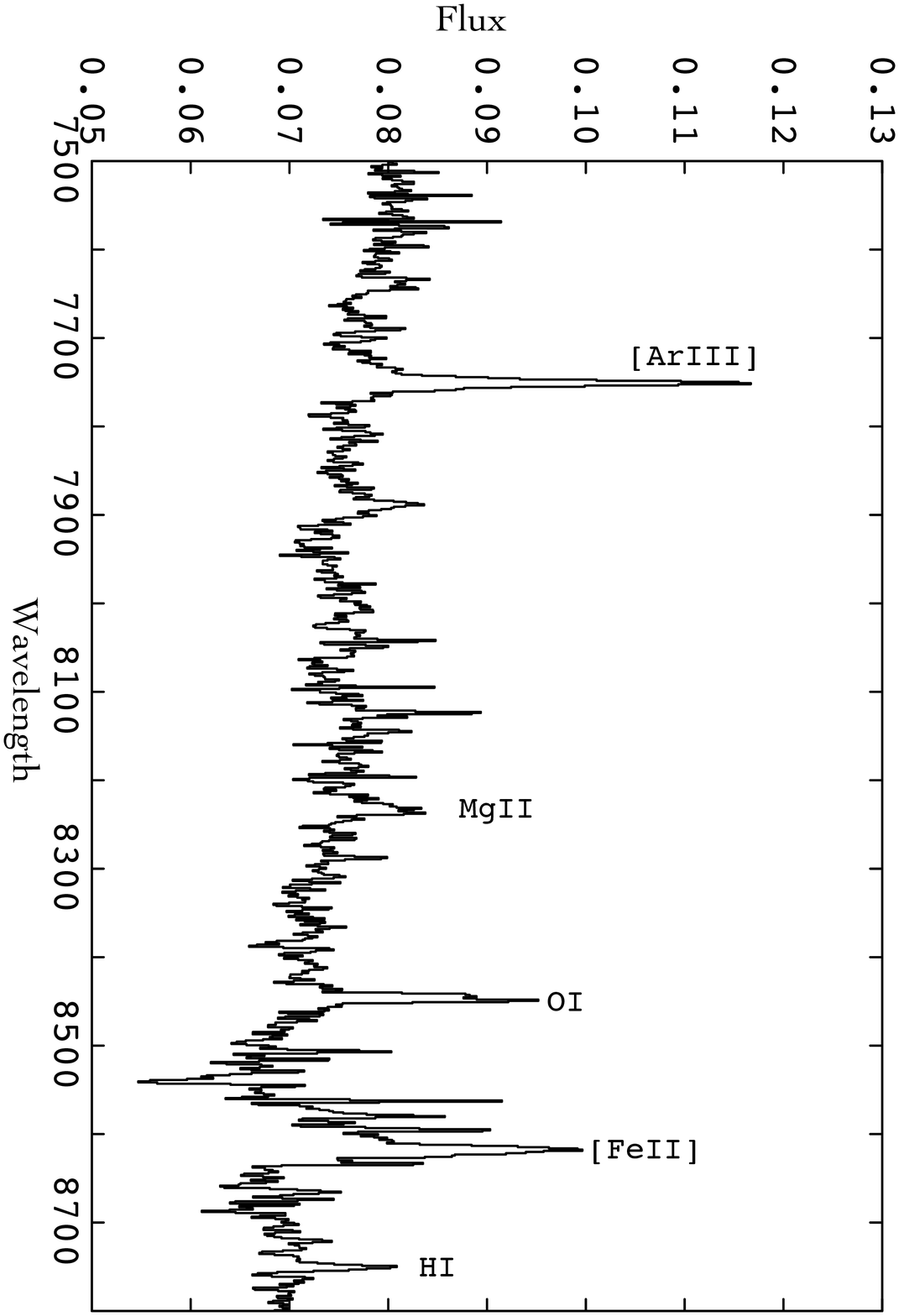}
\vspace{2.7in}
\caption{Four zoomed spectral windows covering the full spectral range are shown to highlight some of the weakest lines.  Unless otherwise specified, fluxes   in these and  other figures are given in  units of    $\times$10$^{-14}$ erg s$^{-1}$ cm$^{-2}$ \AA$^{-1}$. Some relevant lines are indicated.}
\label{fig:spectrab}
\end{figure*}

\subsection{Spectral fitting}

In order to study the kinematic and physical properties of the nuclear ionized outflow we have fitted  the spectral profiles of numerous emission lines (Sect. 3). For this we used the  {\sc STARLINK} package {\sc DIPSO}. This software  is based on the optimization of fit coefficients, in the sense of minimizing the sum of the squares of the deviations of the fit from
the spectrum data. The output from a completed fit consists of the optimized parameters (Gaussian central $\lambda$, FWHM, intensity peak, flux) and their errors (calculated in the linear
approximation, from the error matrix).

Two methods were attempted. In method I,
we assumed  that all emission lines have the same kinematic substructure as [OIII]$\lambda\lambda$4959,5007  (same number of kinematic components with identical FWHM in \kms  and relative velocities, as explained in Villar Mart\'\i n et al. \citeyear{vm14}).  In physical terms, this method corresponds to a situation where all gaseous regions emit all lines, although with different relative fluxes.  

On the other hand, this is not necessarily the case and some lines might come from different regions, possibly resulting in  variations of the kinematic substructure from line to line. As an example,  low critical density lines ($\sim$few$\times$10$^3$ cm$^{-3}$) are quenched in high density gas  with $n>$10$^6$ cm$^{-3}$. To account for this possibility, whenever possible we have also attempted  to fit the lines 
   without applying  prior kinematic restrictions (method II). 

Also, all fits that produced unphysical results were rejected even if mathematically valid (e.g. a fit producing [NII] $\frac{\lambda6583}{\lambda6548}$ very different from the theoretical value 3.0).

As we will see the lines are complex with multiple kinematic components  and they are often severely blended with neighbour lines 
so that the fits are sometimes complicated and not unambiguous. Applying methods I and II whenever possible we account for the uncertainties involved in a more realistic way than using a single method.

Both methods do not always produce acceptable results. For instance, the [SII]$\lambda\lambda$4068,4076+H$\delta$ and 
[SII]$\lambda\lambda$6716,6731 blends could not be successfully fitted applying full constraints from the [OIII] lines. The opposite occurs with the H$\alpha$+[NII] blend. 

Methods I and II could be successfully applied to H$\gamma$, [OIII]$\lambda$4363, H$\beta$ and [OI]$\lambda$6300. It is found that, [OIII]$\lambda$4363 is the most uncertain. While the narrow and intermediate components isolated in the fits (see below) are consistent within 
30 and 10 per cent, respectively, with the two methods, a discrepancy of a factor of $\sim$60 per cent is found for the broadest component. The results for all three kinematic components
isolated in the fits of the other three lines differ by  $<$20 per cent.  These uncertainties will be taken into account when relevant.

\section{Results}

\subsection{Line identification.}

We identify $\sim$90 emission lines  in the SDSS optical spectrum of MRK~477 (Table  \,\ref{tab:tablines}), many of them detected for the first time in this object.  
For some features the identification is ambiguous and/or it could be the contribution of several emission lines. In such cases,  all the possible identifications or contributors are quoted separated with ``;". As an example, a broad feature  with an asymmetric profile is detected  at $\lambda\sim$5272 \AA.
 This is identified as ``[Fe III];[FeII];[FeII];[Fe VII]" because it might be the blend of  		[FeIII]$\lambda$5270.4, [FeII]$\lambda\lambda$5273.4,5276.0 and [FeVII]$\lambda$5276.4. When two or more lines are known to contribute to a given feature, all are quoted and separated with ``+". 
For instance, the [OII] at $\lambda\sim$3727 and [NI] at $\lambda\sim$5200 doublets are shown as ``3727.0+3728.8" and ``5197.9+5200.4".

The works by \cite{ver04} and \cite{ver13} have been used for iron (Fe) line identifications.
The critical densities $n_{crit}$ of  forbidden transitions are quoted in column 5 of Table  \,\ref{tab:tablines}  when available. Most have been computed with the {\sc PYNEB} package
optimized for the analysis of emission lines (Luridiana, Morisset \& Shaw \citeyear{lur14}). These have been estimated for $T_e\sim$15,000 K. The rest have been retrieved from \cite{derob84}.

The lines identified in the NIR spectrum are shown in  Table  \,\ref{tab:tabnearir}.
 
\begin{table*}
\centering
\tiny
\begin{tabular}{lllll}
\hline
(1) & (2) & (3) & (4)  & (5) \\
Species & $\lambda_{\rm air}$ & $\lambda_{\rm obs}$ &   $\frac{Flux}{Flux(\rm H\beta)}$ & $n_ {crit}$\\   
 & (\AA) &   (\AA)  &   &  (cm$^{-3}$) \\ \hline
~[OII]	&  3727.0+3728.8 	& 	3726.9	&  2.26 $\pm$ 0.18 & (1.3/4.5)$\times$10$^3$  \\
~H12  	&  3750.2 		&	3750.1		&  0.014 $\pm$ 0.003 \\
~[FeVII] 	& 3758.9		& 	3759.3		&  0.046 $\pm$ 0.005	& 3.7$\times$10$^7$$^a$  \\
~H11     	&  3770.6	 	& 3770.4  		&  0.023 $\pm$ 0.002 \\
~H10     	& 3797.9		& 3797.5	  	&  0.026 $\pm$ 0.003 \\	
~H9 	& 3835.4		& 3835.5 	&  0.045 $\pm$ 0.003 \\
~[NeIII] 	& 3868.8	&	3868.6	&	1.04 $\pm$ 0.03  & 1.3$\times$10$^7$\\
~HeI+H8 	&  3888.7+3889.05 & 	3888.8	& 0.16 $\pm$ 0.02	 \\
~[NeIII]+H7 & 	3967.5+3970.1 &	3968.2  & 0.43 $\pm$ 0.04	& 1.3$\times$10$^7$   \\
~HeI  	& 4026.2  &	4026.1	& 0.013 $\pm$ 0.002\\ 	
~[SII] 	& 4068.6 	&	4068.6 & 0.19 $\pm$ 0.01 &  	 3.0$\times$10$^6$	 \\
~[SII] 	& 4076.4	&	4076.1  &  0.065 $\pm$ 0.03 &  1.5$\times$10$^6$ \\
~H$\delta$ & 4101.7	 &	4101.5	&  0.24 $\pm$ 0.02 \\
~[FeV] 	& 4180.9	 &	4180.7   &  0.008 $\pm$ 0.002 	\\
~FeII];[FeV]	& 	4227.2;4228.0 & 4228.2$\pm$0.1   &    0.017 $\pm$ 0.002 \\ 
~[FeII]	 & 4244.0  & 	4244.5 & 0.019 $\pm$ 0.002 \\	
~[FeII] 	 & 		4276.8  & 4278.3$\pm$0.1 & 0.013 $\pm$ 0.002  \\
~[FeII] 	 & 		4287.4  &   4289.1$\pm$0.4 &   0.014 $\pm$ 0.002  \\		
~H$\gamma$ &	4340.5 &	4340.31 &  0.45 $\pm$ 0.05\\
~[OIII] &	4363.2  &	4363.2   & 0.23 $\pm$ 0.02  & 3.4$\times$10$^7$\\ 
~[FeII] &  4416.3 & 	4416.4  & 0.021 $\pm$ 0.002   \\	
~HeI	& 4471.4 & 	4472.2 & 0.035 $\pm$ 0.002	 \\
~[FeII]	 & 4570.0  & 	4570.9 & 0.018 $\pm$ 0.005 \\	
~[FeIII];CIV    & 4658.1	 & 4658.7 &  0.034 $\pm$ 0.001 \\
~HeII	& 4685.7 &	4685.7  &   0.138 $\pm$ 0.008\\	
~[FeIII]  	& 4701.5	 & 4700.2 &   0.010 $\pm$ 0.008 \\
~[ArIV]  	& 4711.4 &	4711.5 & 0.011 $\pm$ 0.002 & 1.7$\times$10$^4$	\\
~[NeIV] 	& 4725.5;4726.8	  &  4725.1	& 0.008 $\pm$ 0.003 & (6.2/2.5)$\times$10$^7$	\\
~[ArIV] 	& 4740.2	& 4740.2 &	  0.016 $\pm$ 0.003 & 1.6$\times$10$^5$ \\
~[FeIII]  	& 4754.7 	& 4754.0$\pm$0.5 &  0.007 $\pm$ 0.002\\
~[FeIII]  	&  4769.5 & 4769.6 &    0.006 $\pm$ 0.002 \\ 
~[FeIII]  	&   4777.7 &  4777.5 &   0.007 $\pm$ 0.002 \\ 
~[Fe II]  & 4814.5 &  4816.3$\pm$  &    0.016 $\pm$ 0.002 \\  
~H$\beta$ & 4861.3   &  	4861.3	& 1.00 & \\ 
~[FeVII]  	&	4893.4  &	4893.9 & 0.008 $\pm$ 0.002	&  \\ 
~[OIII] 	& 4958.9  &  4958.9  &   3.7 $\pm$ 0.3 & 8.6$\times$10$^5$\\
~[OIII] 	&  5006.8 &  5006.8 &   10.5 $\pm$ 0.3  & 8.6$\times$10$^5$ \\
~[FeVI]	 &	5145.8  &	5145.9	& 0.007 $\pm$ 0.002 & 2.7$\times$10$^7$$^a$	   \\ 
~[FeVII] 	&	5158.9 &	5159.4	& 0.042 $\pm$ 0.005	& 3.3$\times$10$^6$$^a$    \\
~[FeVI]  	& 	5176.4 &	5177.0	& 0.010 $\pm$ 0.001  & 3.3$\times$10$^7$$^a$	 \\
~[NI]  	& 5197.9+5200.4	& 5199.2 &   0.05 $\pm$ 0.01  & (1.6/0.51)$\times$10$^3$ \\
~[Fe III];[FeII];[FeII];[Fe VII] & 		5270.4;5273.4;5276.0;5276.4	& 	5272.1 &   0.058 $\pm$ 0.004   \\
 ~[CaV]  	& 	5309.1  &	5309.7	&	0.032 $\pm$ 0.002 & 5.1$\times$10$^7$ \\
~[FeVI]  	&	5335.2		& 5335.5	 &	 0.017 $\pm$ 0.002  \\ 
 ~[FeII]   &		5379.0  &    5379.7 	&  0.011 $\pm$ 0.002  \\
 ~HeII;[FeII] 	&    	5411.5;5412.7		& 5412.0 &  0.015 $\pm$ 0.005	& 	\\
~FeII];[FeVI]	&  5425.3;5425.7	& 	5424.1	&   	0.018 $\pm$ 0.006	&  \\
~[FeVI] &  5485.0 &    5484.5  & 0.005 $\pm$ 0.001	 \\
~[Cl III]	&   5517.7		& 	5517.5	&	0.006 $\pm$ 0.002  &  8.5$\times$10$^3$\\  
 ~[ArX];[FeII];FeII]  & 	5533.0;5527.6;5534.8 & 	5531.39	&0.013 $\pm$ 0.003	 &   \\ 
~[Cl III]  	& 5537.9 &	5537.9	 &  0.009 $\pm$ 0.002    &   2.9$\times$10$^4$  \\
~[OI]	& 5577.34 &	5577.9 & 	0.014 $\pm$ 0.001	 &  9.0$\times$10$^7$ \\  
~[FeVI]	 &  5631.1  &	5630.9	&  0.008 $\pm$ 0.002 	\\  
~[FeVI] 	&	5677.0 	 & 5677.1 & 	0.005 $\pm$ 0.001 \\
~[FeVII] 	& 5720.9	& 5721.8$\pm$0.2 &	 0.047 $\pm$ 0.002  &	3.6$\times$10$^7$$^a$ \\
~[NII] 	& 5754.6 &	5754.7  &	0.040 $\pm$ 0.002 & 1.8$\times$10$^7$	  \\
~HeI & 5875.6 & 		5875.7 & 0.131 $\pm$ 0.007	\\	
~[FeVII]  	& 	6086.9 &	6087.7 & 0.12 $\pm$ 0.01	& 3.6$\times$10$^7$$^a$  \\
~[OI] 		&	6300.3 &	6300.5 & 0.81 $\pm$ 0.02	&  1.1$\times$10$^6$   	  \\
~[SIII]  	& 6312.1 	&	6311.8	&	0.016 $\pm$ 0.003  &  1.6$\times$10$^7$  \\ 
~[OI]  	&  6363.8 & 	6365.0	 &	0.26 $\pm$ 0.02  & 1.1$\times$10$^6$ \\
~[ArV] 	&       6435.1 & 	6435.7	&0.010 $\pm$ 0.002	 & 1.4$\times$10$^7$ \\
~[NII]  	& 	6548.1  & 	6548.1  &	0.53 $\pm$ 0.05  & 1.2$\times$10$^5$	  \\
~H$\alpha$ & 	6562.8  & 	6562.8	 & 3.9 $\pm$ 0.2 \\ 	
~[NII]  	&	6583.5 & 	6583.4	& 1.5 $\pm$ 0.1	 & 1.2$\times$10$^5$  \\
~HeI  	& 	6678.2 & 	6678.7   &  0.035 $\pm$ 0.005   \\		
~[SII] 	&	6716.4  &	6716.5	& 0.60 $\pm$ 0.03 &	  1.5$\times$10$^3$\\
~[SII]		&      6730.8 &	6730.9	& 0.75 $\pm$ 0.03 	& 4.0$\times$10$^3$\\
~[ArV]  	&	7005.8	 & 7005.7	&	0.014 $\pm$ 0.003 & 1.3$\times$10$^7$  \\
~HeI		&       7065.7 &	7065.3	&0.07 $\pm$ 0.005	 \\
~[ArIII] 	&	7135.8	 & 7135.8	& 	 0.20 $\pm$ 0.01 &  6.4$\times$10$^6$\\
~[FeII] 	&	7155.2	 	& 7156.8		& 0.06 $\pm$ 0.01  \\
~[FeII] &	7172	.0 		& 7173.8 	& 0.014 $\pm$ 0.004			 \\
~[CaII] 	& 	7291.5 	& 7292.8 &  0.02 $\pm$ 0.01 	\\
~[OII]		& 	7318.9 	&	7321.0  & 0.23 $\pm$ 0.04 & 6.5$\times$10$^6$		\\
~[OII]		& 	7339.7	& 7329.2	 &	0.32 $\pm$ 0.04 & 6.5$\times$10$^6$  \\
~[NiII]		& 	7377.8 	&	7379.1	& 0.040 $\pm$ 0.008  \\
~[FeII]  &  7452.6 & 		7453.9$\pm$0.5   &    0.016 $\pm$ 0.002 \\
~FeI;[FeII]	&	7634.0;7637.5	 &   7639$\pm$2	&  0.007 $\pm$ 0.002\\
~[ArIII]  	&	7751.1	 &	7750.9	&	0.062 $\pm$ 0.003 &  6.4$\times$10$^6$\\
~MgII; [NiIII]; [FeXI]; MgII   &	7877.0; 7889.9; 7891.8;7896.0 	& 7888.8	 &0.023 $\pm$ 0.003	 \\
~MgII & 8232.0 & 8232.7$\pm$0.5 & 0.020 $\pm$ 0.005 \\
~OI 		&	8446.0 	&	8446.6	& 0.029 $\pm$ 0.002 	 \\
~HI Pa 14		&  8598.5	 	&	 8595.4		& 0.010 $\pm$ 0.003\\
~[FeII] 	& 	8605.5 	& 	8604.5 	&	0.033 $\pm$ 0.004	&	  \\ 
~[FeII] &	8617.0  & 8617.8 &   0.05 $\pm$ 0.02 \\ 
~HI Pa 12  & 8750.6 &  8750.2 & 	0.013 $\pm$	0.002	\\
\hline
\end{tabular}
\caption{Emission lines identified in the SDSS spectrum of MRK~477. The observed line wavelengths $\lambda^{\rm obs}$ (column 3) have been determined by fitting a single Gaussian to the emission lines. Errors are quoted when they provide relevant information regarding possible multiple identifications or there is a significant shift between the observed and the air wavelengths (column 2). The line ratios (column 4) are given relative to H$\beta$, with $F(\rm H\beta)$=(1.01$\pm$0.04)$\times$10$^{-13}$ erg s$^{-1}$ cm$^{-2}$. They are not corrected for reddening. The critical densities are quoted when available. Superscript $^a$ in column (5) indicates that $n_{crit}$ has been taken from De Robertis \& Osterbrock (1984). The other $n_{crit}$ values have been computed with 
the {\sc PYNEB} software (see the text).}   
\label{tab:tablines}
\end{table*}

\begin{table*}
\centering

\begin{tabular}{lllll}
\hline
(1) & (2) & (3) & (4)  & (5) \\
Species & $\lambda_{\rm air}$ & $\lambda_{\rm obs}$ &   $\frac{Flux}{Flux(Br\gamma)}$ & $n_ {crit}$\\   
 & (\AA) &   (\AA)  &   &  (cm$^{-3}$) \\ \hline
~[Si X]	& 1.4305	& 	1.4308	&  0.61$\pm$0.10 &   6.3$\times$10$^8$$^{(a)}$  \\
~[Fe II]  &  1.5339 & 1.5337	&  0.79$\pm$0.10 &  4.6$\times$10$^{4}$$^{(b)}$  \\
~[Fe II]  &  1.6440 & 1.6435 	& 4.5$\pm$0.5   &  5.6$\times$10$^{4}$$^{(b)}$ \\
~Pa$\alpha$  &  1.8756  & 1.8746 	& 12.6$\pm$1.4  &   \\
~H$_2$ 1-0S(3)  &  1.9576 & 1.9564	&  0.84$\pm$0.10 &    \\
~[Si VI]	& 1.9629	& 	1.9626	&  0.97$\pm$0.12  &   6.3$\times$10$^8$$^{(a)}$ \\
~H$_2$ 1-0S(2)  &  2.0338 & 2.0329	&  0.20$\pm$0.02  &   \\
~HeI & 2.0587 & 2.0575 & 0.29$\pm$0.03 \\
~H$_2$ 1-0S(1)  &  2.1218 & 2.1206 	&   0.63$\pm$0.17&   \\
~Br$\gamma$  &  2.1661 &  2.1650	&   1.00$\pm$0.15 &   \\
\hline
\end{tabular}
\caption{Emission lines identified in the  NIR spectrum of   of MRK~477 extracted from a $\sim$1 arcsec $\times$ 3.5 arcsec~ aperture centered on the nucleus. $^{(a)}n_{\rm crit}$ from Rodr\'\i guez Ardila et al. (2011). 
$^{(b)}n_{\rm crit}$ from Drougados et al. (2010).
Fluxes are quoted relative to Br$\gamma$, for which F(Br$\gamma$)=(3.0$\pm$0.3)$\times$10$^{-15}$ erg s$^{-1}$ cm$^{-2}$, although the absolute flux calibration is  uncertain (see the text).}   
\label{tab:tabnearir}
\end{table*}

\begin{figure}
\includegraphics{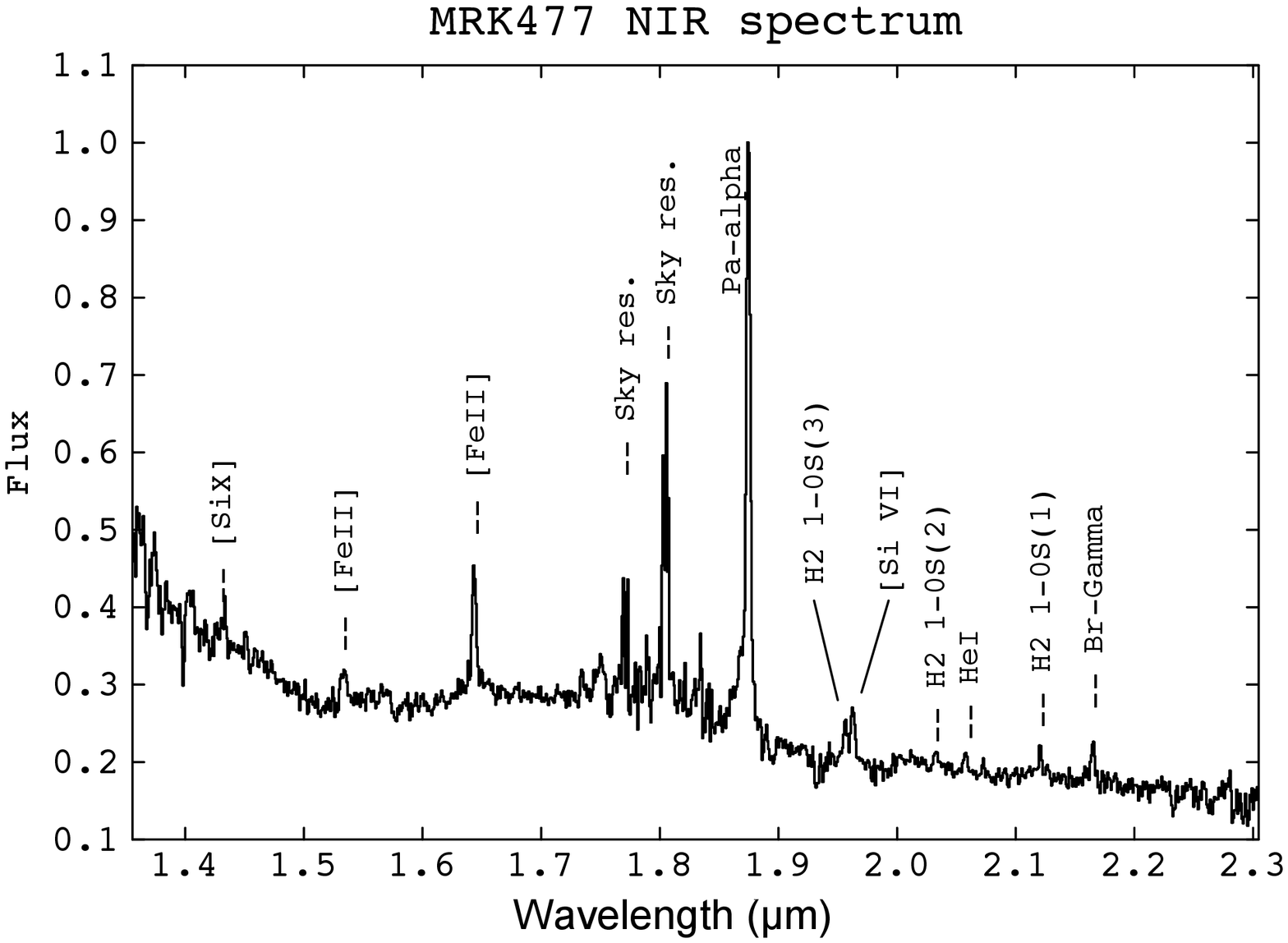}
\vspace{2.5in}
\caption{NIR LIRIS-WHT spectrum of MRK~477 extracted from a $\sim$1\arcsec$\times$3.5\arcsec~ aperture centered on the nucleus. The detected lines are indicated.  Flux is 
in units of $\times$10$^{-15}$ erg s$^{-1}$ cm$^{-2}$ A$^{-1}$ and $\lambda$ in $\mu$m.} 
\label{fig:liris}
\end{figure}

\begin{table}
\centering
\tiny
\begin{tabular}{llll}
\hline
Comp. &  FWHM & $V_S$ & $\frac{F_i}{F(\rm \rm H\beta)}$ \\ 
&	(km s$^{-1}$) & (km s$^{-1}$) \\  \hline
& &  [SII]$\lambda$4068  &   \\  \hline
 Narrow &  94$\pm$37 &  -2$\pm$8 & 	0.07$\pm$0.01  \\
   Interm. &    505$\pm$35 &  17$\pm$10 & 	0.26$\pm$0.02  \\
Broad &      1534$\pm$97 &  -376$\pm$85 &	0.25$\pm$0.05  \\ \hline 
& &  [SII]$\lambda$4076  &   \\  \hline
 Narrow &   94$\pm$37 & -2$\pm$8 & 	0.010$\pm$0.006  \\
   Interm. & 505$\pm$35 & 17$\pm$10 & 	0.12$\pm$0.02  \\
Broad &    1534$\pm$97 & -376$\pm$85 &	 0.08$\pm$0.07  \\ \hline 
& &  H$\delta$ &  \\  \hline
 Narrow &  94$\pm$37 & -2$\pm$8 & 	0.23$\pm$0.04  \\
   Interm. &  505$\pm$35 & 17$\pm$10 & 	0.23$\pm$0.06  \\
Broad  & 1534$\pm$97 & -376$\pm$85 &	0.26$\pm$0.03  \\ \hline 
& &   H$\gamma$ &   \\  \hline
 Narrow &  157$\pm$28 & -7$\pm$7 &	0.47$\pm$0.03 \\
   Interm. &   529$\pm$40 & 36$\pm$10 & 	0.38$\pm$0.04  \\
 Broad 	&  1830$\pm$133 &	-224$\pm$51 & 0.57$\pm$0.06  \\  \hline
& &  [OIII]$\lambda$4363  &   \\  \hline
   Narrow & 157 $\pm$28	& -7$\pm$7  & 0.11$\pm$0.02  \\
   Interm. &  529$\pm$40	 &  36$\pm$10  & 0.43$\pm$0.03  \\
 Broad &  1830$\pm$133 	& -224$\pm$51 	 & 0.27$\pm$0.08  \\  \hline
& &    H$\beta$ &   \\  \hline
     Narrow &  137$\pm$22 & -2$\pm$6 &	(4.1$\pm$0.4)$\times$10$^{-14}$  \\
   Interm. &  471$\pm$22 & 7$\pm$7 & 	(3.7$\pm$0.1)$\times$10$^{-14}$  \\
 Broad & 1527$\pm$75 &	-188$\pm$24 & (2.3$\times$0.1)$\times$10$^{-14}$  \\  \hline
& &  [OIII]$\lambda$5007  &  \\  \hline
 Narrow    & 	177$\pm$15 & 	5$\pm$9 &     8.5$\pm$0.9 \\
   Interm. & 545$\pm$15 & 22$\pm$11 &	13.1$\pm$0.5 \\
 Broad 	& 1839$\pm$53 & -227$\pm$23 &	9.6$\pm$0.7  \\  \hline
 & & [OI]$\lambda$6300 &  \\  \hline
Narrow & 211$\pm$11 &  -4$\pm$5 & 	0.60$\pm$0.07 \\
Interm. &  563$\pm$29 &	17$\pm$6 & 0.91$\pm$0.08 \\ 
Broad & 1745$\pm$40 &	 -257$\pm$24 & 1.08$\pm$0.07 \\   \hline
& & H$\alpha$ & \\  \hline
Narrow &   As [OIII] & As [OIII] &  3.1$\pm$0.3 \\
Interm. &  As [OIII]   &	As [OIII]  & 4.9$\pm$0.2 \\
 Broad & As [OIII]  & As [OIII] 	& 4.2$\pm$0.3 \\  \hline
& & [NII]$\lambda$6583 & \\  \hline
Narrow &  As [OIII]  & As [OIII]  & 1.0$\pm$0.1 \\
 Interm. &  As [OIII]  & As [OIII]  & 1.70$\pm$0.07 \\
Broad & 	As [OIII]  & As [OIII] & 0.9$\pm$0.2\\   \hline
& & [SII]$\lambda$6716  &  \\  \hline
Broad & 186$\pm$9  & -3$\pm$5 & 	0.67$\pm$0.07  \\
   Interm. &501$\pm$17 & -4$\pm$6 &	0.68$\pm$0.04  \\
 Broad & 1397$\pm$76 & -493$\pm$89 &	0.30$\pm$0.08  \\  \hline
& & [SII]$\lambda$6731 &   \\  \hline
 Narrow &  186$\pm$9  &	    -3$\pm$5 & 	  0.66$\pm$0.07  \\
Interm. &   501$\pm$17 &  -4$\pm$6 &		0.80$\pm$0.03  \\
Broad &   1397$\pm$76 &  -493$\pm$89  &	0.77$\pm$0.08 \\   \hline
 \end{tabular}
\caption{Results of the spectral decomposition of the main lines in the spectrum of MRK~477. Three kinematic components are isolated in all cases, including a broad blueshifted component emitted by the ionized outflow.  No values are quoted for H$\alpha$ and [NII]
because successful fits could only be obtained with full constraints from the [OIII] lines (see Sect. 2.1).
 The  flux of each component  relative to   H$\beta$ is given in the last column. The exception is H$\beta$,   for which the actual fluxes are quoted.}  
\label{tab:tabfits}
\end{table}

\subsection{Relationships between  the kinematics and $n_{crit}$, IP$_{\rm low}$ and IP$_{\rm high}$.}

We plot in  Fig. \ref{fig:kin-corr}  the FWHM (top panels) and the velocity shift $V_S$  (bottom) versus  the critical density ($n_{crit}$, Table  \,\ref{tab:tablines}), the lower (IP$_{\rm low}$) and upper 
(IP$_{\rm high}$) ionization potentials for a subsample of 
 $\sim$28 optical and 2 NIR forbidden  emission lines.\footnote{The FWHM and $V_S$ values have been derived from single Gaussian fits to the emission lines.} $V_S$ corresponds to  the difference 
between $\lambda_{\rm air}$ and $\lambda_{\rm obs}$ in Table  \,\ref{tab:tablines},  where $\lambda_{\rm air}$  is the vacuum wavelength and $\lambda_{\rm obs}$ is that measured from the spectrum corrected for $z$.   Only lines for which both parameters could be measured with reasonable accuracy are plotted. As an example, this was not the case for [SiX]$\lambda$1.430 and [FeII]$\lambda$1.257 due to the weakness of the lines and/or the distortion of the line profiles by sky residuals}.  The optical and NIR lines span a wide range in critical densities  ($\sim$1.3$\times$10$^3$-6.3$\times$10$^8$ cm$^{-3}$) and ionization potentials (IP$_{\rm high}\sim$13-205 eV).
  A significant correlation is found between the FWHM  and the critical density (Spearman correlation coefficients $r_s$=0.79 and $p$=0.000001, excluding [FeII]$\lambda$1.644), as already found by De Robertis \& Osterbrock (1986)
whose data are shown as open circles. No significant correlation is found  with the ionization potentials ($r_s$=0.43 and $p$=0.02 for FWHM versus  IP$_{\rm low}$
and $r_s$=0.42 and $p$=0.03 for FWHM versus  IP$_{\rm high}$). 
 No trends are found either for $V_S$ except maybe an apparent preference for redshifts ($V_S>$0)  for lines with the highest critical densities.

The NIR [FeII]$\lambda$1.644 line is a clear outlier in Fig. \ref{fig:kin-corr}, being too broad compared to what is expected for its critical density (5.6$\times$10$^4$  cm$^{-3}$) and ionization potentials (7.9 and	16.2 eV) according to the general trends defined by the other lines. We will discuss this in more detail below.

\begin{figure*}
\includegraphics{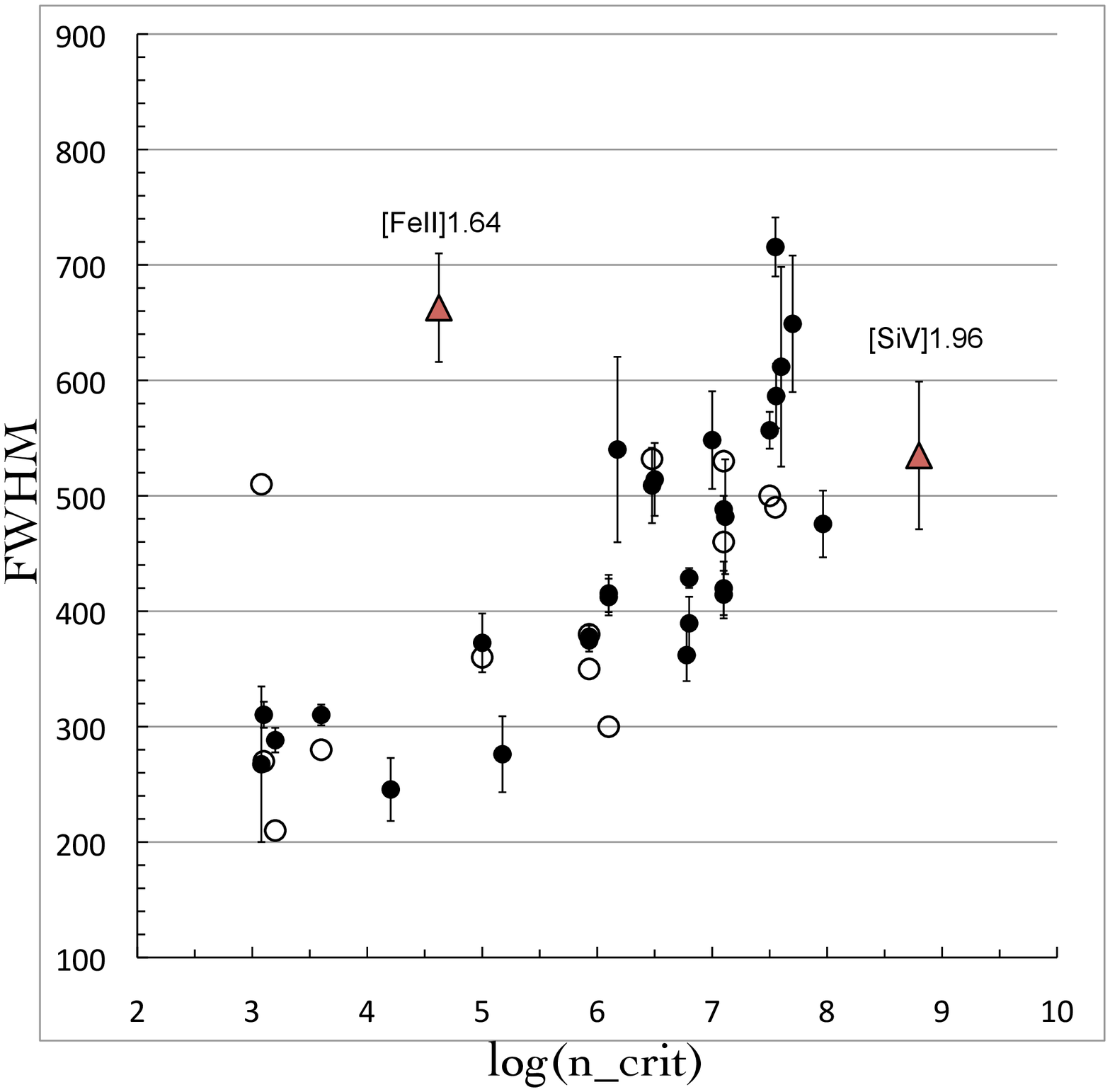}
\includegraphics{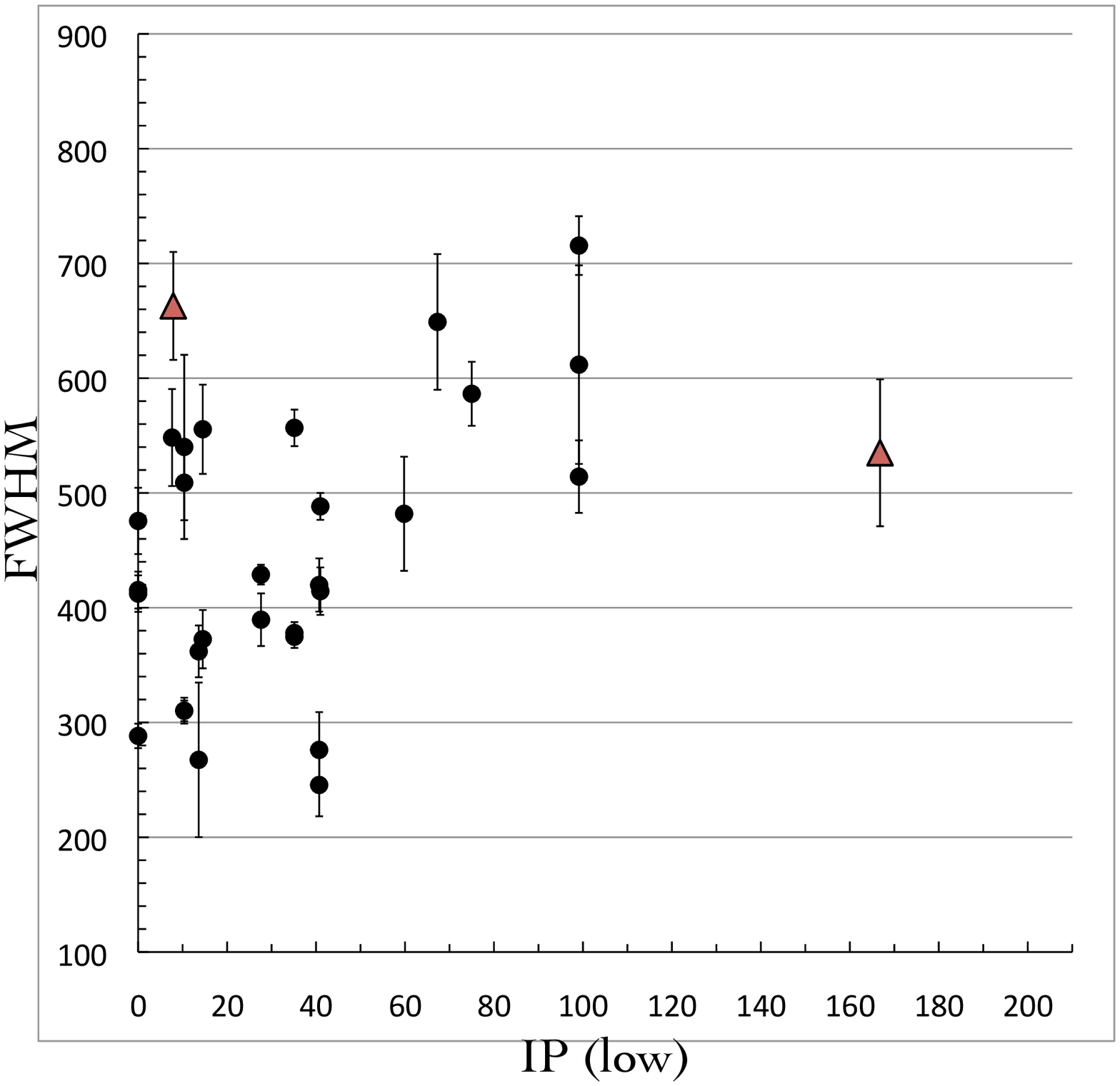}
\includegraphics{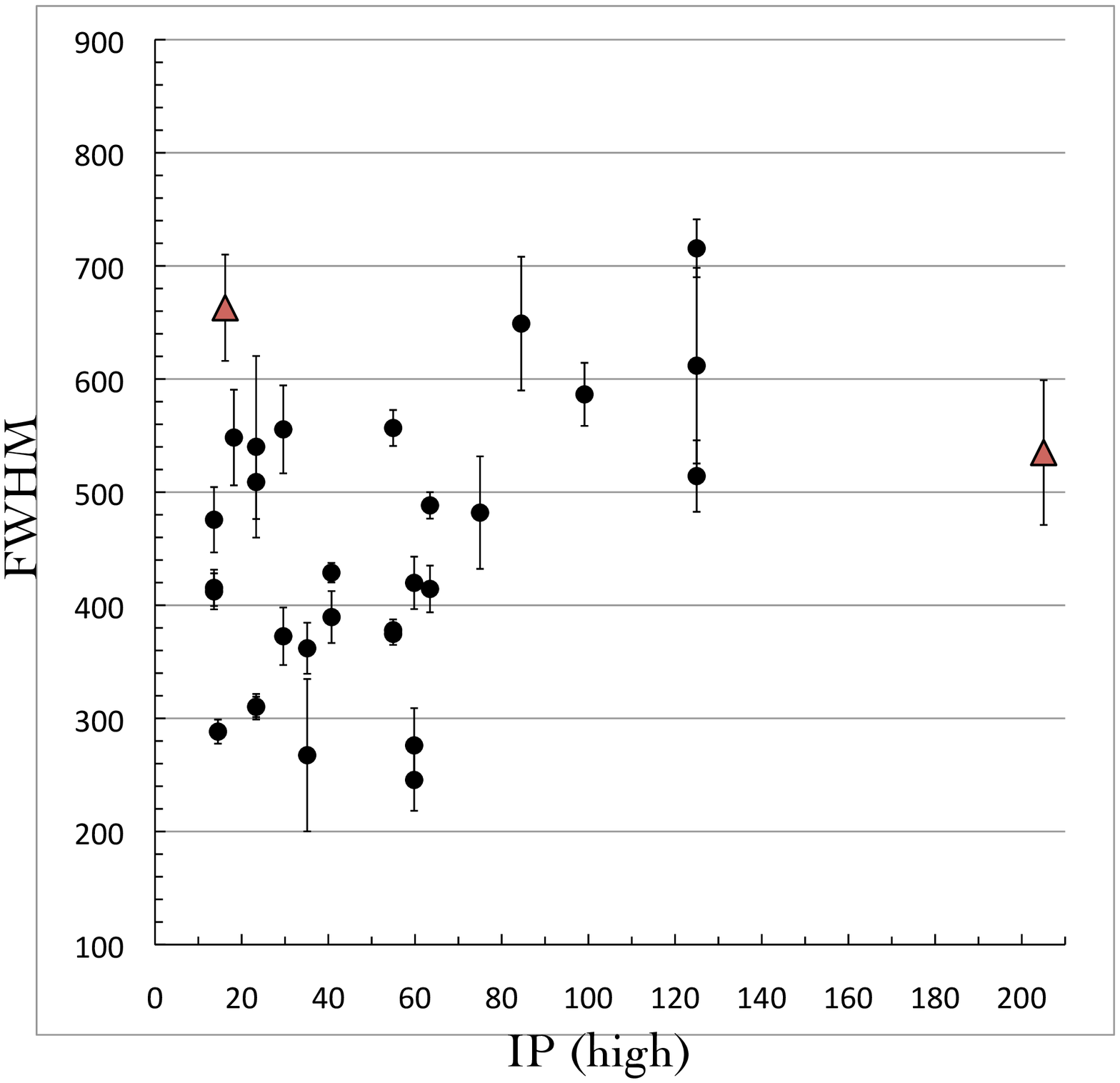}
\vspace{2.5in}
\includegraphics{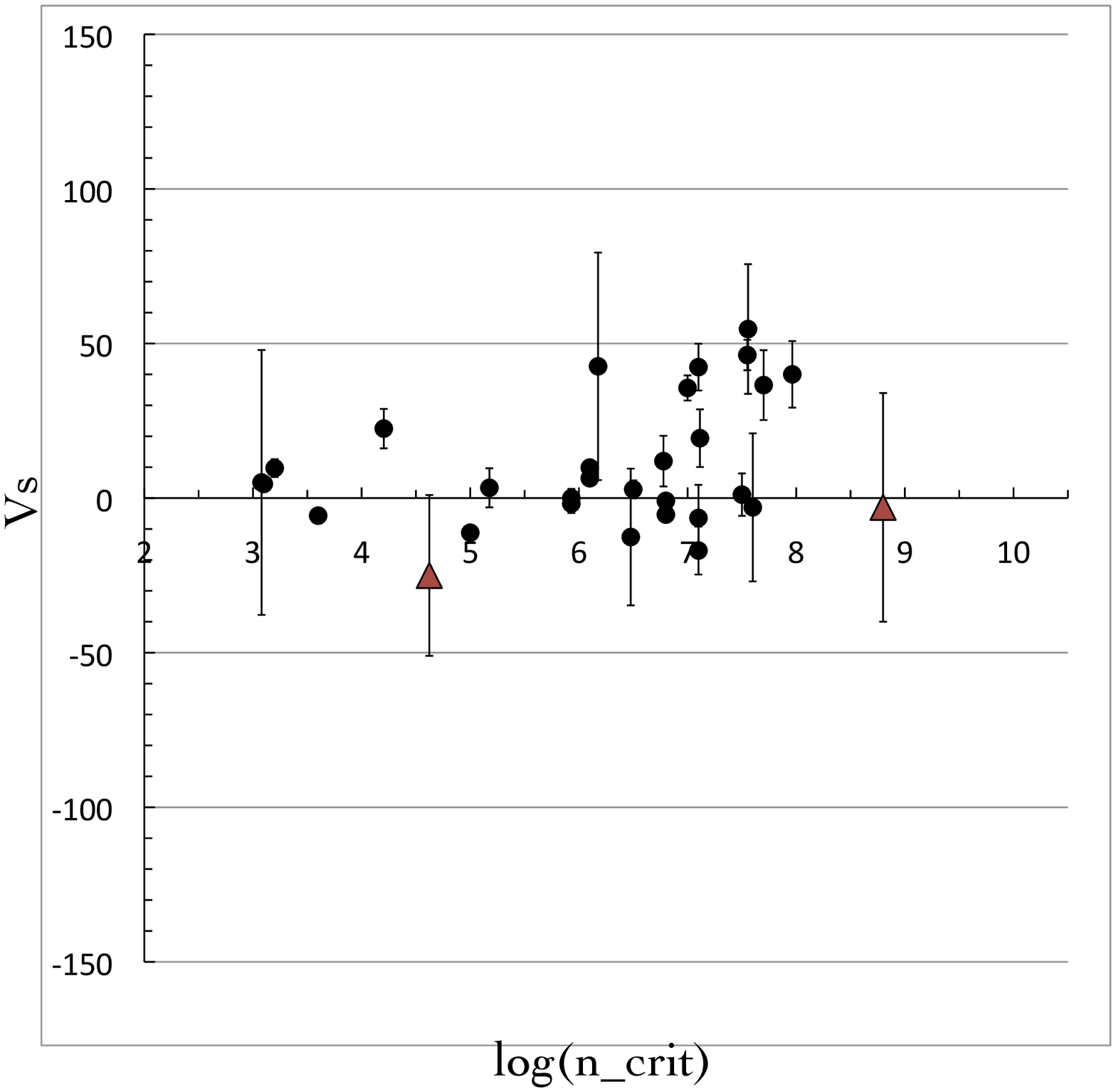}
\includegraphics{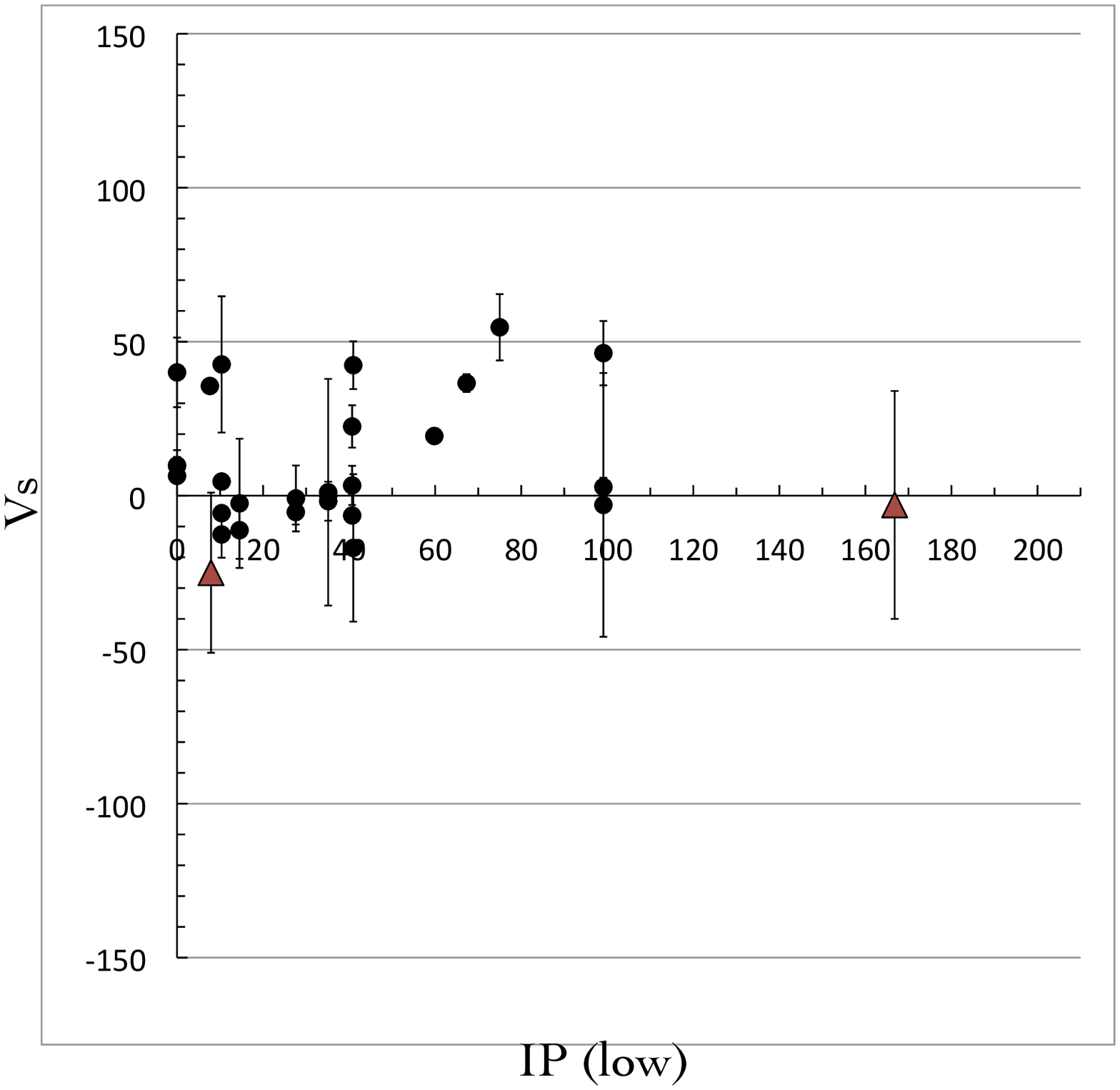}
\includegraphics{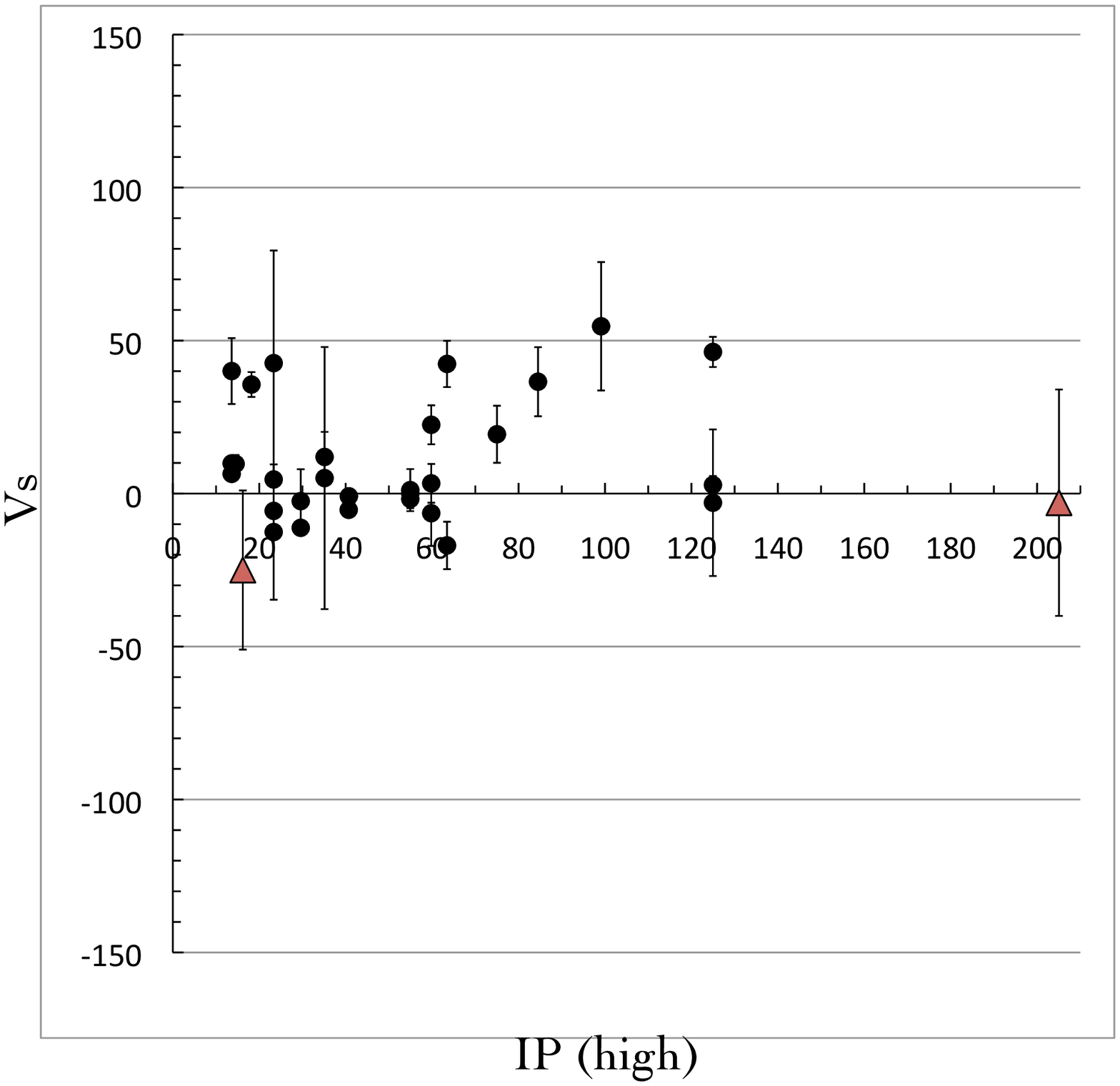}
\vspace{2.5in}
\caption{The FWHM (top panels) and the velocity shift $V_s$ (bottom panels) of   a subsample of $\sim$30 forbidden lines are plotted against
the critical density (log($n_{\rm crit}$)), the lower and higher ionization potentials of the species involved.
The measurements from De Robertis \& Osterbrock (\citeyear{derob86}, dR \& Ost 86) are also shown in the top-left panel as open circles.
The outlier at log($n_{\rm crit})\sim$3.1 and FWHM$\sim$500 km s$^{-1}$ corresponds to  [OII]$\lambda$3727, which appears artificially  broader due to blended doublet components. The NIR lines are marked with red triangles.
[FeII]$\lambda$1.644   is a clear outlier in the top panels. It is too broad compared to what is expected from the general trend defined by the rest of the lines.}
\label{fig:kin-corr}
\end{figure*}

\begin{figure}
\includegraphics{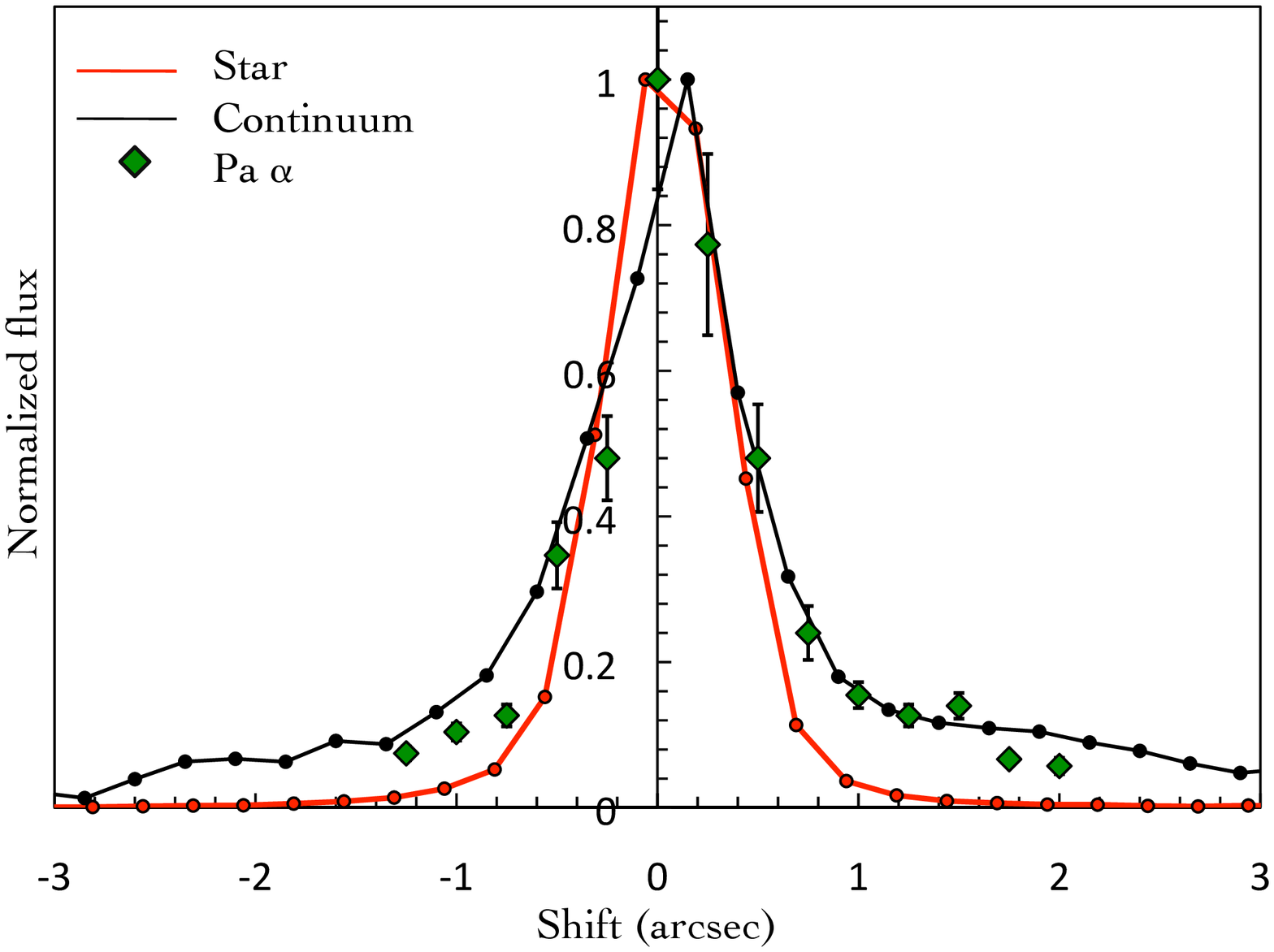}
\vspace{2.5in}
\caption{Spatial profiles (normalized fluxes) of the NIR continuum and Pa$\alpha$ along PA 43$\degr$ N to E, compared with the seeing
disk (star). Both the continuum and Pa$\alpha$ are spatially extended.}
\label{fig:spatPa}
\end{figure}

The correlation between FWHM and $n_{crit}$  has been  found in many type 1 and type 2 AGNs and it holds for AGN  luminosities  that differ by a factor of up to $\sim$5000  (Espey \citeyear{esp94}).
$V_S$, on the other hand, shows a different behavior from object to object (e.g. Appenzeller \& \"Ostreicher \citeyear{app88}). 

The FWHM versus  $n_{crit}$ correlation implies   that the line emission from the NLR originates in different subregions
under different physical  conditions and kinematic properties  with a broad density range. 
Higher critical density lines (many of which have also very high ionization potentials) are broader because they 
are emitted predominantly by higher velocity,  high density and high ionization regions closer to the central engine.  

We have done $n$ estimations using the density-diagnostic line ratios  [SII]$\lambda$6716/$\lambda$6731, [Cl~III]$\lambda$5517/$\lambda$5537 and [ArIV]$\lambda$4711/$\lambda$4740, which are not affected by  extinction or  $T_e$ variations. They are sensitive to different $n$ regimes.   The   [SII] ratio ($n_{crit}$=1.5$\times$10$^3$ and 4.0$\times$10$^3$  cm$^{-3}$, respectively) is sensitive to $n$ variations $\sim$several$\times$10$^3$ cm$^{-3}$. These lines are suppressed in high density regions by collisional de-excitation. Other line pairs that can be overcome this limitation are
 [Cl III]$\lambda$5517/$\lambda$5537 ($n_{crit}$=8.5$\times$10$^3$ and 2.9$\times$10$^4$ cm$^{-3}$), sensitive at  $n\sim$10$^4$ cm$^{-3}$ and [ArIV]$\lambda$4711/$\lambda$4740 ($n_{crit}$= 1.7$\times$10$^4$ and 1.6$\times$10$^5$ cm $^{-3}$) at $n\sim$10$^4$-10$^5$ cm$^{-3}$.   
 
For MRK~477,  [SII]$\lambda$6716/$\lambda$6731=0.70$\pm$0.06, [ClIII]$\lambda$5517/$\lambda$5537=0.74$\pm$0.22  and [ArIV]$\lambda$4711/$\lambda$4740 =0.67$\pm$0.09,  impling densities    $n$=1945$^{+670}_{-460}$ cm$^{-3}$ (from [SII]),   7250$^{+13,300}_{-5100}$ (from  [Cl III]) and $n$=14,700$^{+6150}_{-3750}$  (from [ArIV]). So, the [SII] doublet implies the existence of gas with $n\sim$2000 cm$^{-3}$, the 
[Cl III] doublet, in spite of the larger uncertainties suggests  higher densities and  the [ArIV] doublet confirms the existence of gas with $n>$10$^4$ cm$^{-3}$ in the NLR of MRK~477. 

\subsection{Spatial extension of the ionized gas}

The only available information about the spatial distribution of the ionized gas in MRK~477 is provided by 
the  [OIII] image presented by \cite{hec97}. It shows a bright knot or ridge of emission at $\sim$0.4 arcsec to
the northeast of the central source and aligned with the radio axis. This knot is not resolved from the nucleus in our NIR spectrum, for  which the slit  
 was roughly aligned along  the same axis. We find, on the other hand, that Pa$\alpha$  is extended at both sides of the central source. 
We show in Fig. \,\ref{fig:spatPa} the spatial profiles of (1) a star observed during the May run (seeing size FWHM=0.65$\pm$0.05),
(2) the  MRK 477 continuum and (3) the P$\alpha$ flux, with the underlying continuum subtracted. Although more compact than the continuum,
Pa$\alpha$ is extended, as demonstrated by the clear excess above the seeing wings,  up to a maximum radial distance of $\sim$2 arcsec  or $\sim$1.5 kpc to the NE.

\begin{figure*}
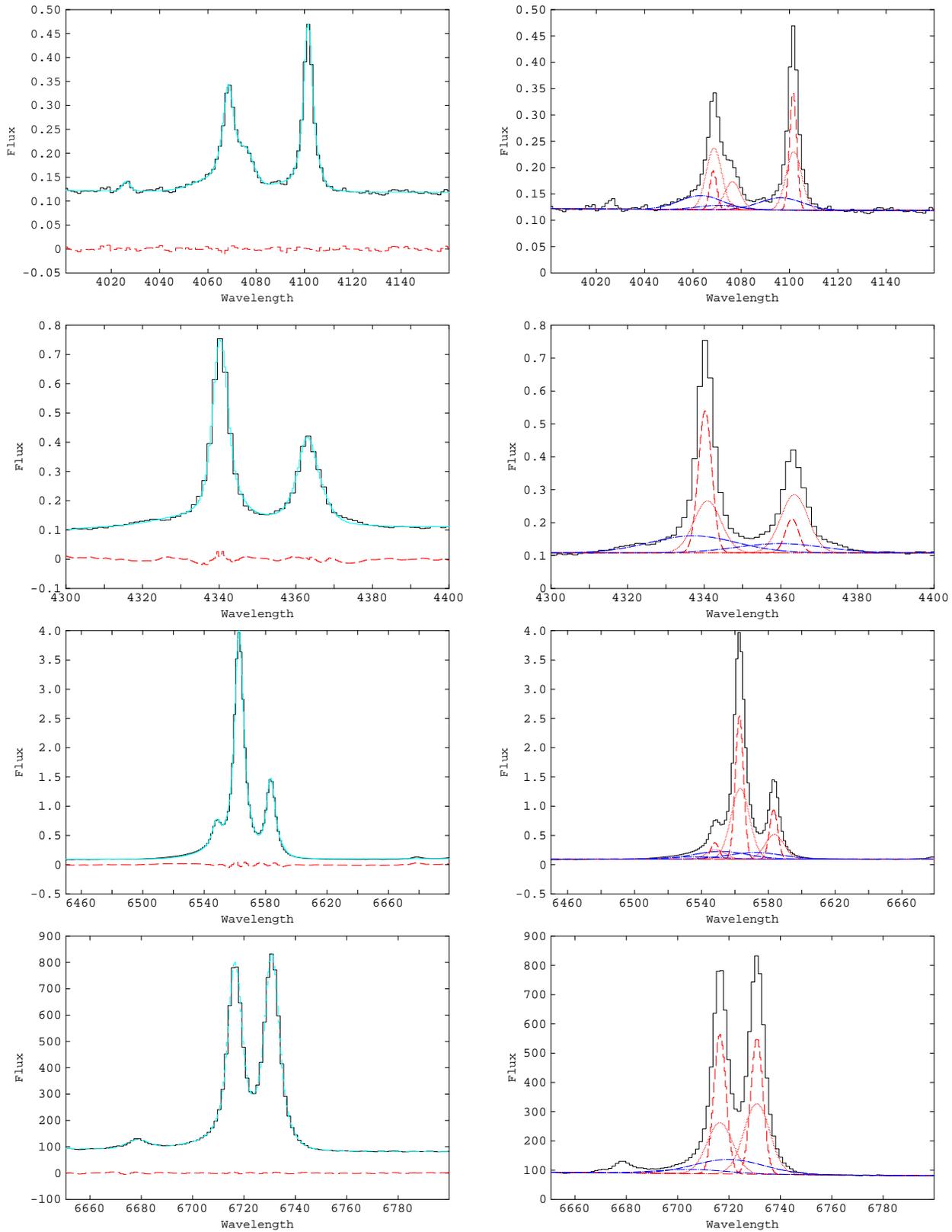

\includegraphics{fitsiihd-1440-residuals-new.ps}
\includegraphics{fitsiihd-1440-i-new.ps}
\vspace{2.1in}
\includegraphics{fitoiiitemp-1440-residuals-new.ps}
\includegraphics{fitoiiitemp-1440-new.ps}
\vspace{2.1in}
\includegraphics{fitniiha-residuals-new.ps}
\includegraphics{fitniiha-i-new.ps}
\vspace{2.1in}
\includegraphics{fitsiired-residuals.ps}
\includegraphics{fitsiired-i.ps}
\vspace{2.1in}
\caption{Left: data (black),  fits (cyan)  and residuals (dashed-red) for a diversity of emission lines. Right: individual kinematic components isolated in each line.  Different line  styles  are used for different  kinematic components. The same red colour  is used for components with the same redshift:  long-dashed red (narrow component), dotted-red (intermediate component), dot-dashed blue (broad component; i.e. the outflow emission).
From top to bottom:  [SII]$\lambda\lambda$4068,4076 and H$\delta$;  H$\gamma$ and [OIII]$\lambda$4363; [NII]$\lambda\lambda$6548,6583 and H$\alpha$;  [SII]$\lambda\lambda$6716,6731.} 
\label{fig:fits}
\end{figure*}

\begin{figure*}
\includegraphics{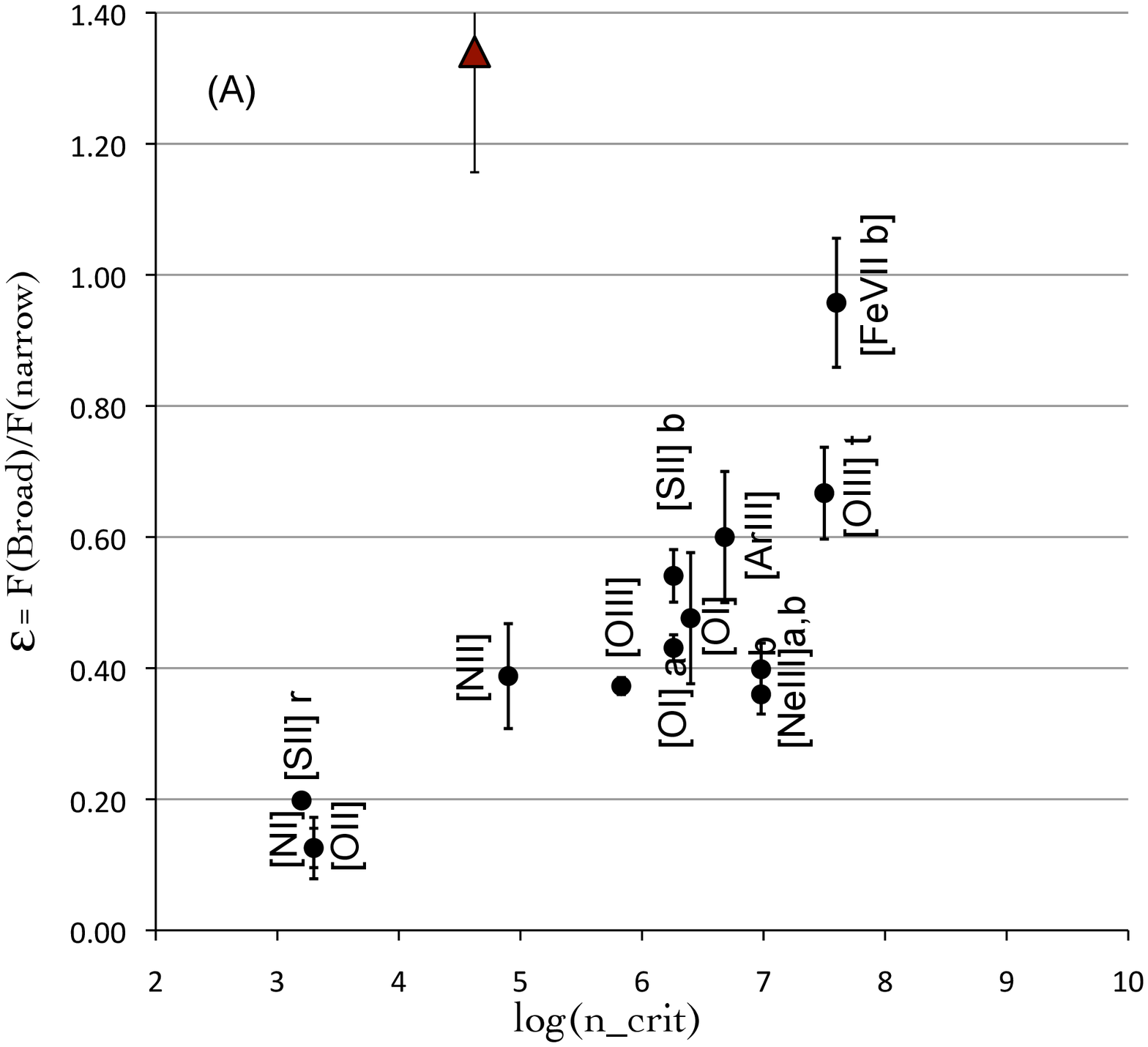}
\includegraphics{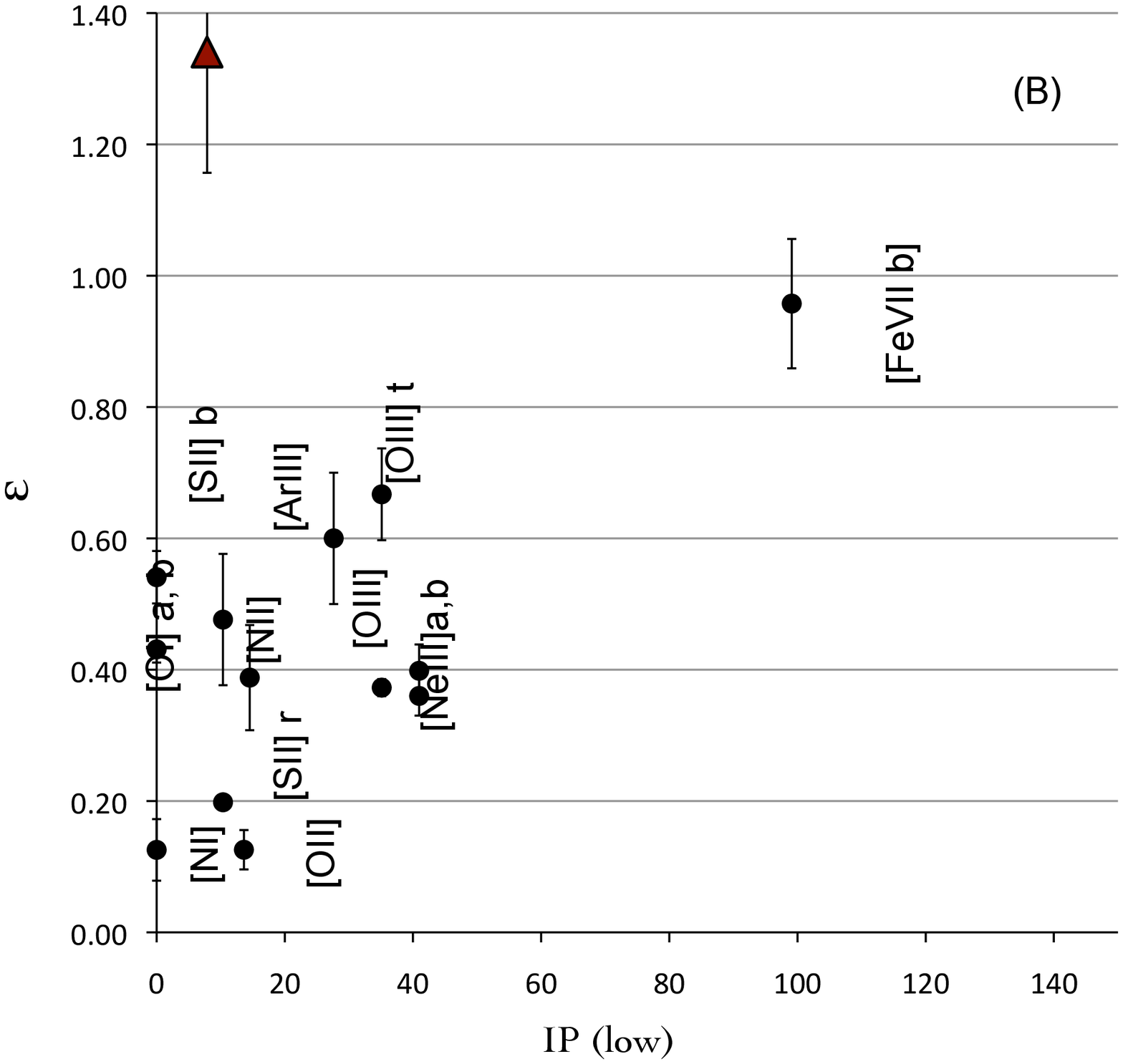}
\includegraphics{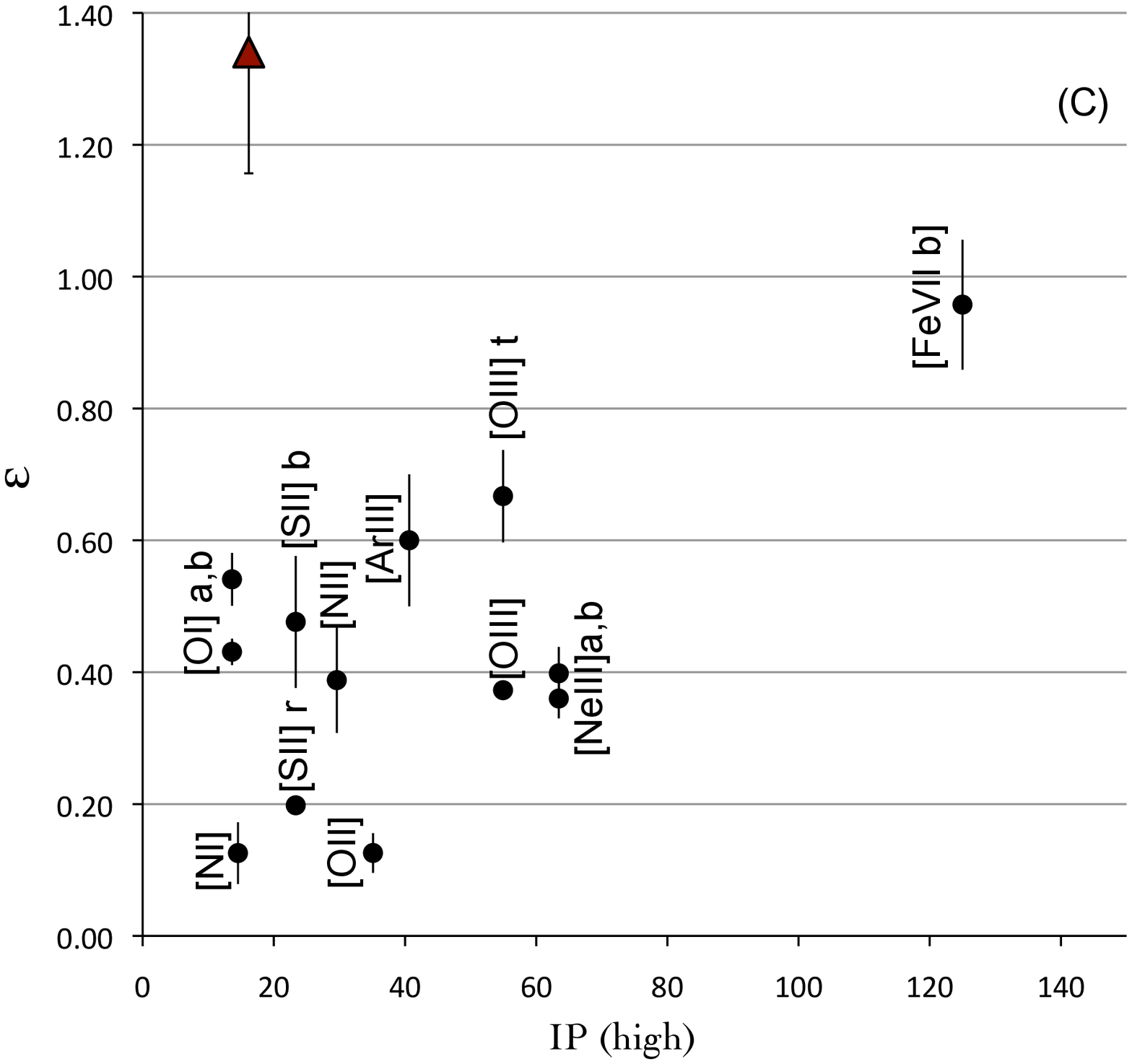}
\vspace{2.6in}
\caption{$\epsilon=\frac{F_{\rm broad}}{F_{\rm narrow}}$ parametrizes the relative contribution between the outflow flux emission and the rest of the line flux. The  correlation with the   critical density (panel A)  implies that the outflow emission is relatively stronger in higher density  gas. 
[OI] `a'  and `b'  correspond to the  $\lambda$6300 and $\lambda$6312 lines respectively;  [NeIII]`a'  and `b'  to $\lambda$3869 and $\lambda$3967;  [SII] `b'  and `r'  refer to the blue and red [SII] doublets at $\lambda\lambda$6048,6076 and $\lambda \lambda$6716,6731 respectively. [FeII]$\lambda$1.644$\mu$m (red triangle) is again an outlier.}
\label{fig:epsilon}
\end{figure*}

\subsection{The ionized outflow}

 \cite{vm14} performed a kinematic and ionization characterization of 
 the nuclear ionized outflow in MRK~477. Its signature is a broad blueshifted kinematic component  in the [OIII]$\lambda\lambda$4959,5007 lines,
with  FWHM$\sim$1850  \kms ~and $V_S\sim$-225 \kms.  Based on the high turbulence of this gaseous component,  the line ratios consistent with AGN (rather than stellar)  excitation processes and the relatively high contribution of the outflowing gas to the total line fluxes,   the authors proposed that
it has been triggered by the 1.2 arcsec   scale radio source. This is supported by the correlation between the radio and [OIII] morphologies (Heckman et al. 1997), which  demonstrates that the radio source is interacting with the NLR.  

\subsubsection{Isolating the outflow emission in a diversity of emission lines}

By means of the  kinematic decomposition of the spectral profiles, we have isolated  the emission from the outflowing gas in numerous emission
 lines  (Sect. 2.1). Our goal is to characterize its physical  properties  and to constrain more accurately its  spatial location.

We show in  Table  \,\ref{tab:tabfits} the results of the fits for several optical emission lines (see also Fig. \,\ref{fig:fits}). 
FWHM and $V_S$ correspond to those fits where no prior constraints from [OIII] were applied (method II, see Sect 2.3).  When both methods
could be applied, the flux values and errors account for the dispersion allowed by them.

Generally, we find rather coherent results for all strong lines for which  multiple component fitting procedures
could be applied. By ``coherent" it is meant that 
 all  lines consist of three kinematic components (Table \,\ref{tab:tabfits}): two of them have similar $z$  and are relatively narrow
 (FWHM$\sim$[95,210] and [470,560] \kms respectively). A third broad blueshifted
 component is moreover isolated in all lines, which is emitted by the outflowing gas. It has FWHM$\sim$[1400,1840 km s$^{-1}$ and 
 $V_S\sim$[-490,-190] \kms. All the three kinematic components have line ratios consistent with type 2  AGN, 
 as already pointed out by  \cite{vm14}.

Other lines show clear evidence of a broad underlying component, although the triple Gaussian fit is not possible  due to the low signal-to-noise and/or the complex blend with neighbour lines.   We have calculated  the ratio  $\epsilon = \frac{F_{\rm broad}}{F_{\rm narrow}}$ between the flux of the broad (outflowing) component and the rest of the line flux  (which for simplicity we will name $F_{\rm narrow}$, although it contains the narrow and intermediate components) for as many lines as possible. This gives a measurement of the relative contribution between the most turbulent outflowing gas and the more quiescent ambient gas.   

The ratio $\epsilon$ is plotted against
log($n_{crit}$), FWHM, IP$_{\rm low}$ and IP$_{\rm high}$  in Fig. \ref{fig:epsilon}.
A significant correlation is found with the critical density ($r_s$=0.73 and $p$=0.003; panel A).  No significant trend  is found with the ionization potentials 
($r_s$=0.40 and $p$=0.15 for $\epsilon$ versus  IP$_{\rm low}$ and $r_s$=0.35 and $p$=0.23 for $\epsilon$ versus  IP$_{\rm high}$).

\begin{figure}
\includegraphics{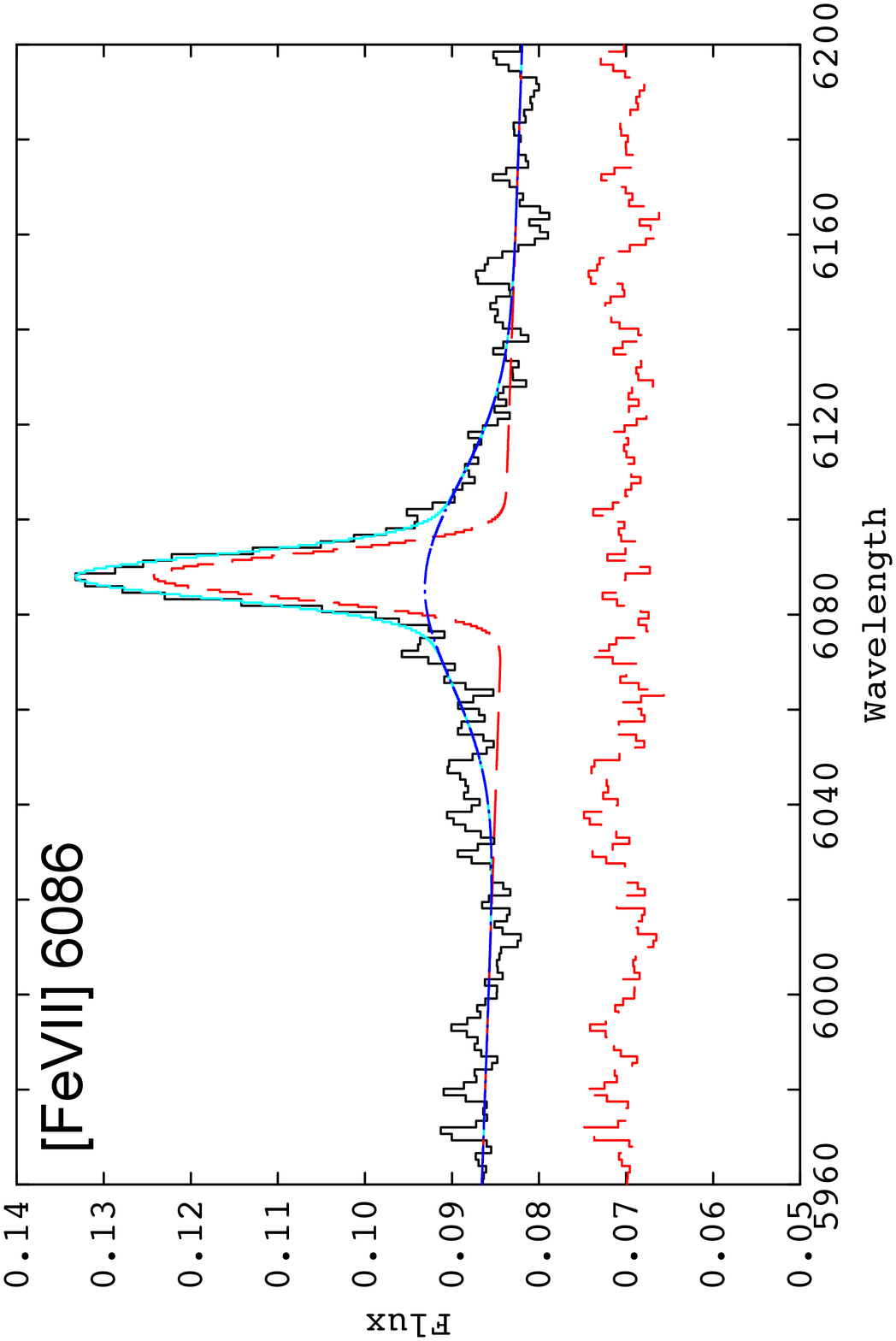}
\vspace{2.5in}
\caption{Fit of the [FeVII]$\lambda$6087 spectral profile. The presence of a very broad underlying wing  (FWHM$\sim$2500 km s$^{-1}$) shows that the high ionization Fe$^{+6}$ region also participates in the outflow. Flux in units of 10$^{-14}$ erg s$^{-1}$ cm$^{-2}$ \AA$^{-1}$.}
\label{fig:fevii}
\includegraphics{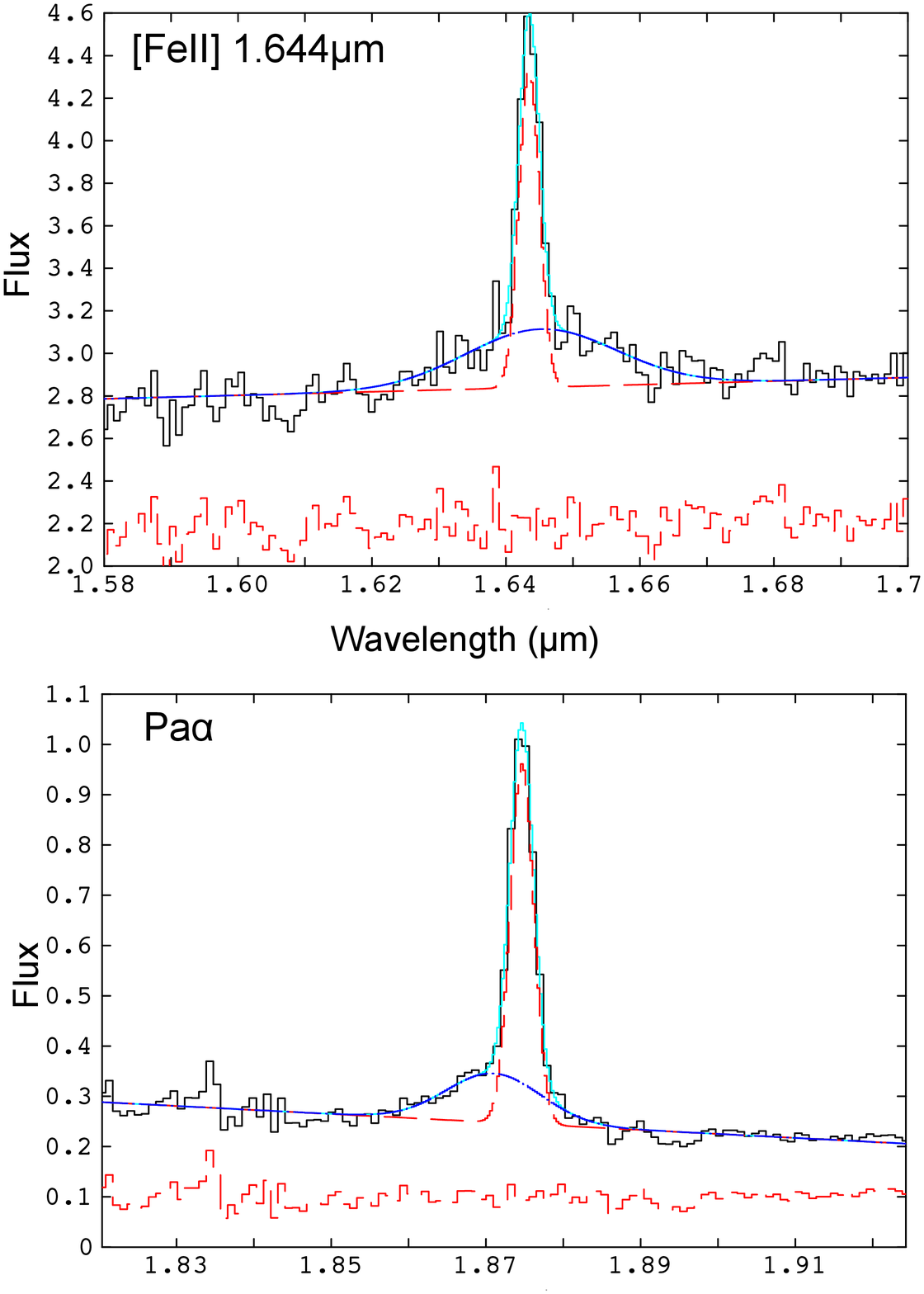}
\vspace{5in}
\caption{Fit of the NIR [FeII]$\lambda$1.644$\mu$m (top) and Pa$\alpha$ (bottom) spectral profiles. Flux in units of 10$^{-16}$ (top) and 10$^{-15}$ (bottom) erg s$^{-1}$ cm$^{-2}$
\AA$^{-1}$.}
\label{fig:fitsNIR}
\end{figure}

An interesting case is the high ionization line [FeVII]$\lambda$6087, for which a  very broad component is isolated    with FWHM=2460$\pm$340 km s$^{-1}$ (Fig. \ref{fig:fevii}) and $V_S$=-150$\pm$60 km s$^{-1}$. 
Thus, the line which shows simultaneously very high critical density and ionization potential, also shows the most  extreme kinematics  and the largest 
$\epsilon$ = 0.96$\pm$0.10 of all optical lines.  At the other end, the   lowest $n_{crit}$ (also low ionization) lines [NI]$\lambda$5200 and [SII]$\lambda\lambda$6716,6731, are the narrowest (FWHM$\sim$300 km s$^{-1}$) and have the smallest $\epsilon\sim$0.1-0.2 (Fig. \ref{fig:epsilon}).

 The correlation between $\epsilon$ and the FWHM  suggests  that the increasing broadening of the lines    is due to the increasing outflow influence. The correlation with  $n_{crit}$ shows that  the outflow emission is relatively stronger in higher density  gas.  
  Ultimately, these results suggest that the FWHM versus  $n_{crit}$ correlation is produced by the outflow in MRK~477.

 [FeII]$\lambda$1.644 is the only forbidden NIR line for which the spectral decomposition could be applied. It consists of a narrow, spectrally unresolved component with FWHM$\la$340 km s$^{-1}$ and a very broad  and prominent underling component with FWHM=4770$\pm$830 km s$^{-1}$ 
 (Fig. \ref{fig:fitsNIR}) shifted by 315$\pm$231 km s$^{-1}$ relative to the narrow core.
It is the only line for which the broad component is not blueshifted. For comparison, Pa$\alpha$ consists of two components with FWHM$\la$260 km s$^{-1}$ and 2240$\pm$230 km  s$^{-1}$ respectively with the broad one blueshifted by -570$\pm$119 km s$^{-1}$.   As in Fig. \ref{fig:kin-corr}, [FeII]$\lambda$1.64$\mu$m is a clear outlier in  all  $\epsilon$ diagrams, with $\epsilon$=1.3$\pm$0.2, the highest of all lines. The line, therefore, has a dominant contribution of emission from the outflowing gas. The different behavior of this line will be discussed later (see below).

\subsubsection{Reddening and electron density}

(i) {\it Reddening correction}. Reddening correction ($c$ and E(B-V)) has been estimated using the Balmer ratios $\frac{\rm H\alpha}{\rm H\beta}$, $\frac{\rm H\gamma}{\rm H\beta}$ 
and $\frac{\rm H\delta}{\rm H\beta}$, for which we assume case B values 2.8, 0.47 and 0.26 respectively, appropriate for
gas densities $n\sim$10$^2$-10$^6$ cm$^{-3}$  and electron temperatures $T_e\sim$10,000-20,000 K (Osterbrock  \citeyear{ost89}). 

We show in Table  \,\ref{tab:reddening} the  E(B-V) and $c$ values derived for the three kinematic components and
the total line fluxes using the expressions:

$$E(B-V) = 1.99 \times \rm log\Big[\frac{{(\rm H\alpha/\rm H\beta)}_{\rm obs}}{2.86}\Big]  ~~~~~~~~~~\rm[eq. 1]$$

and

$$\Big[\frac{F}{\rm H\beta}\Big]_{\rm int}  =  \Big[\frac{F}{\rm H\beta}\Big]_{\rm obs} \times 10^{c ~ [f(\lambda) - f(\rm H\beta)]} ~~~~~~~~~~\rm[eq. 2]$$

where ``$int$''  and ``$obs$''  denote the intrinsic and observed line ratios and
 $f(\lambda) - f(\rm H\beta)$ is given by the standard interstellar extinction curve (Osterbrock \citeyear{ost89}). 
 The final values and errors take 
 into account that negative $c$ values are not allowed\footnote{This is the reason why in Table 3 the same line ratio with different errors (e.g. $\frac{\rm H\gamma}{\rm H\beta}$=0.23$\pm$0.04 and 0.23$\pm$0.06)  can be associated  with slightly different 
$c$ and E(B-V) values.}.  Once  $c$ is determined, other line ratios are
corrected for reddening  using the  equation (2).

\begin{figure*}
\includegraphics{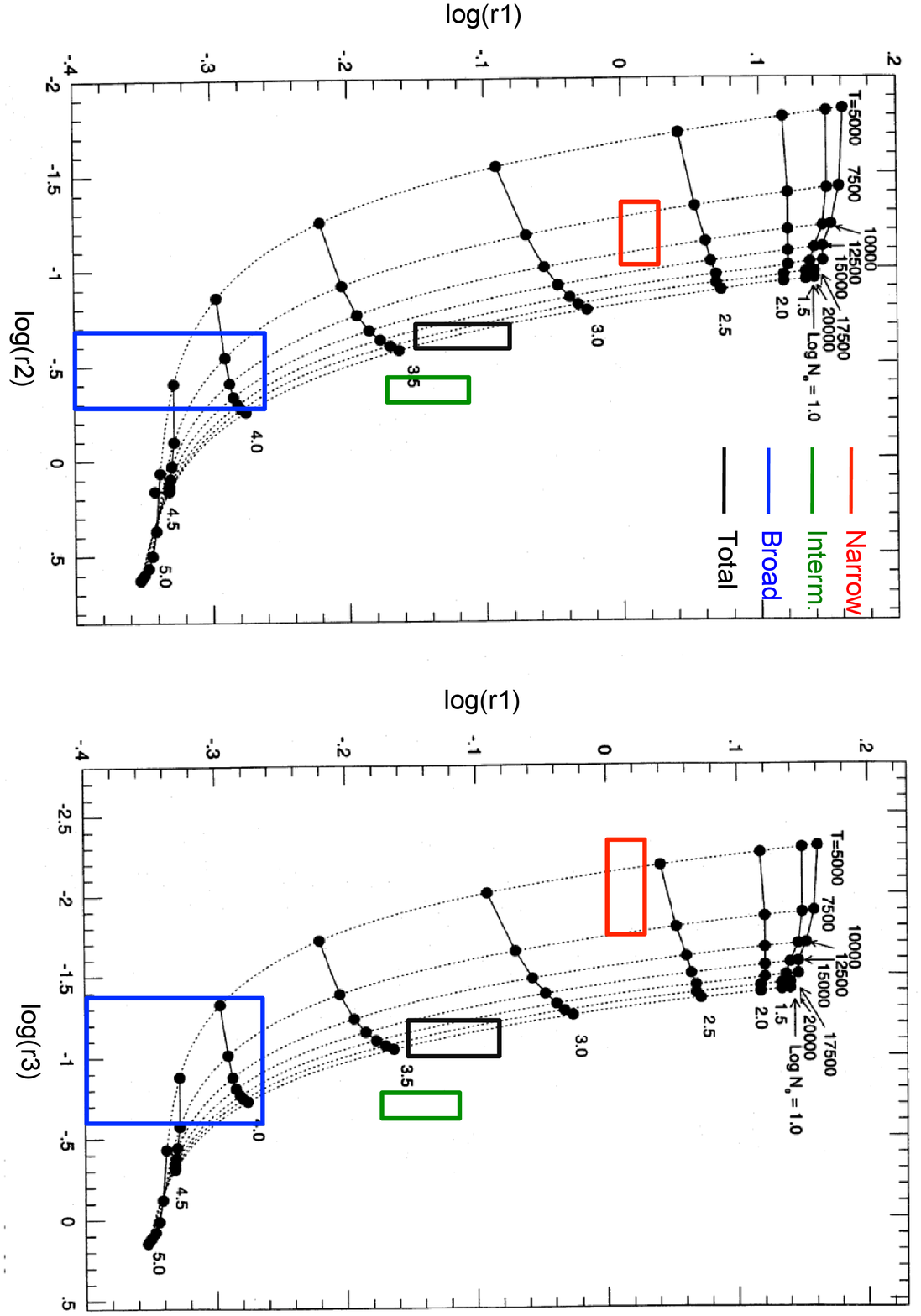}
\vspace{3.4in}
\caption{Constraints on the electron density $n$  for the three kinematic components and for the integrated lines (adapted from Fig. 2 and 3 in Keenan et al. 1996)). A trend is found such that the broader the lines, the higher the density. Thus, the outflowing gas has the highest density.}
\label{fig:keenan}
\end{figure*}

 In spite of the complexity of the line profiles, the reddening values derived from the three ratios are in reasonable good agreement.  From Table 3 we conclude that the narrowest component shows little or no reddening while 
the intermediate component shows the highest reddening. 
For the broad component, the results are less clear, since the Balmer decrement $\frac{\rm H\alpha}{\rm H\beta}$ suggests 
higher reddening (E(B-V)=0.32$\pm$0.06) than the H$\delta$ and H$\gamma$ ratios, which imply little or no reddening. We cannot discard problems with the
H$\alpha$ fit, which is severely blended with the [NII] doublet. However, it might be a real effect. As we will see next,
the broad component is expected to have high densities $n\ga$8000 cm$^{-3}$, maybe up to $\ga$10$^{6}$ cm$^{-3}$, given the detection of very high critical density strong lines. 
Under these circumstances, the Balmer decrement can be enhanced due to collisional excitation of H$\alpha$. $n>$5$\times$10$^{5}$
cm$^{-3}$ will produce $\frac{\rm H\alpha}{\rm H\beta}>4$ (e.g. Binette et al. \citeyear{bin93}). Thus, the inconsistency between the reddening values inferred for the broad component might actually indicate the existence of  high densities in the outflowing gas.

We measure Pa$\alpha$/Br$\gamma$=12.6$\pm$1.4  from the NIR spectrum. Taking errors into account, this implies no reddening or, at most,  or E(B-V)$\le$0.32.
This is consistent with the range of values allowed by the optical decrements.  

\begin{figure}
\includegraphics{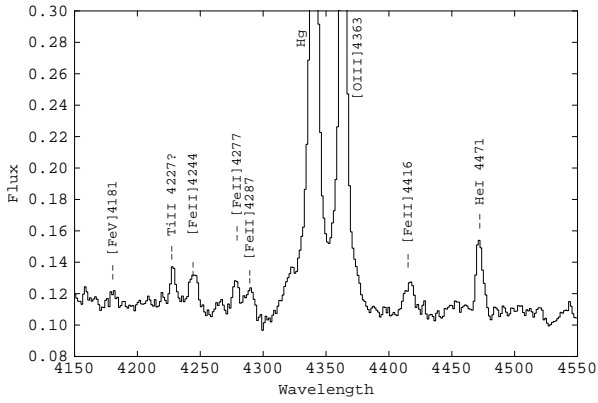}
\vspace{2.2in}
\includegraphics{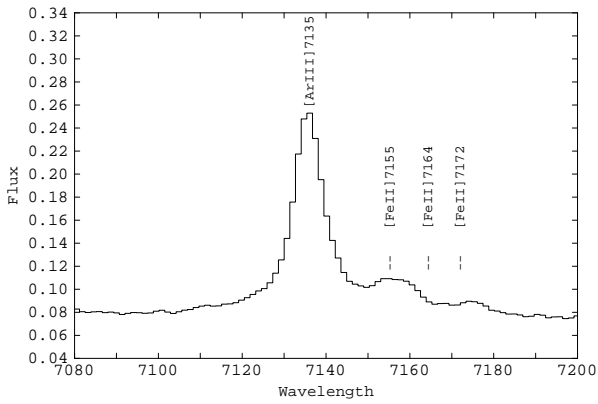}
\vspace{2.2in}
\includegraphics{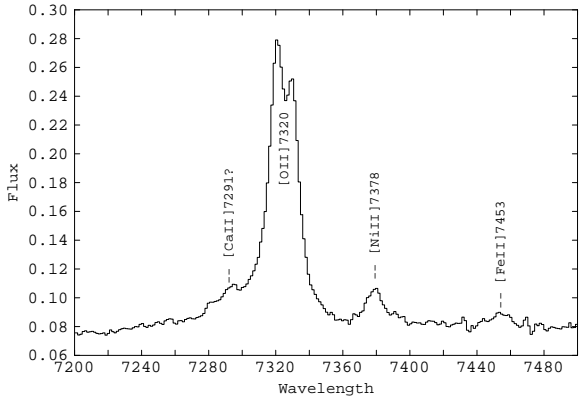}
\vspace{2.2in}
\includegraphics{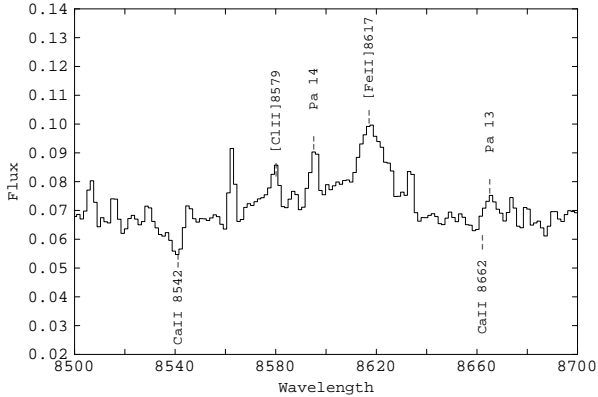}
\vspace{2.2in}
\caption{Several narrow Fe$^+$ emission lines are detected in the spectrum of MRK~477. [CaII]$\lambda$7291 is also tentatively detected (third panel).}
\label{fig:iron2}
\end{figure}

\begin{figure}
\includegraphics{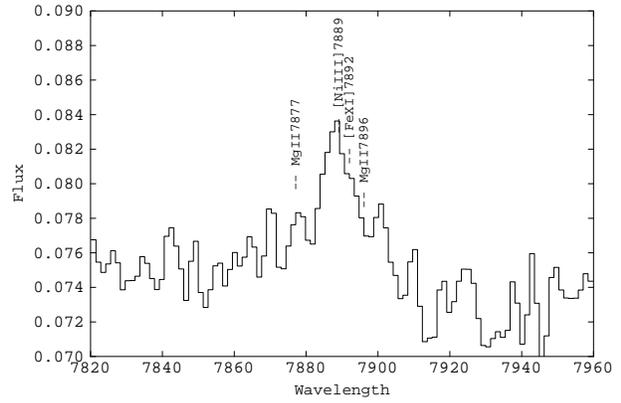}
\vspace{2.2in}
\includegraphics{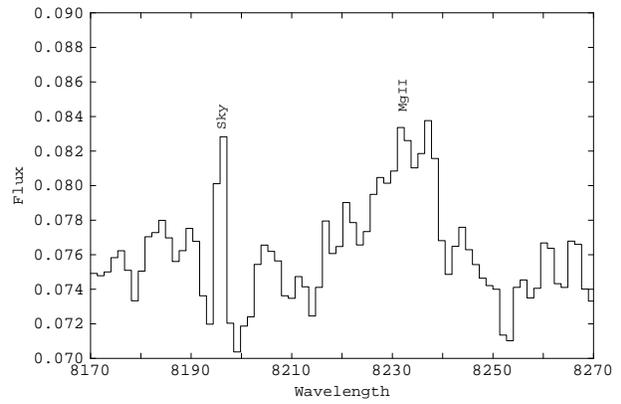}
\vspace{2.2in}
\caption{Emission lines are detected from other  refractory elements apart  from Ca and Fe  (Fig. 9), such as possibly   Ni  (top) and Mg  (top and bottom).}
\label{fig:mag}
\end{figure}

\begin{figure}
\includegraphics{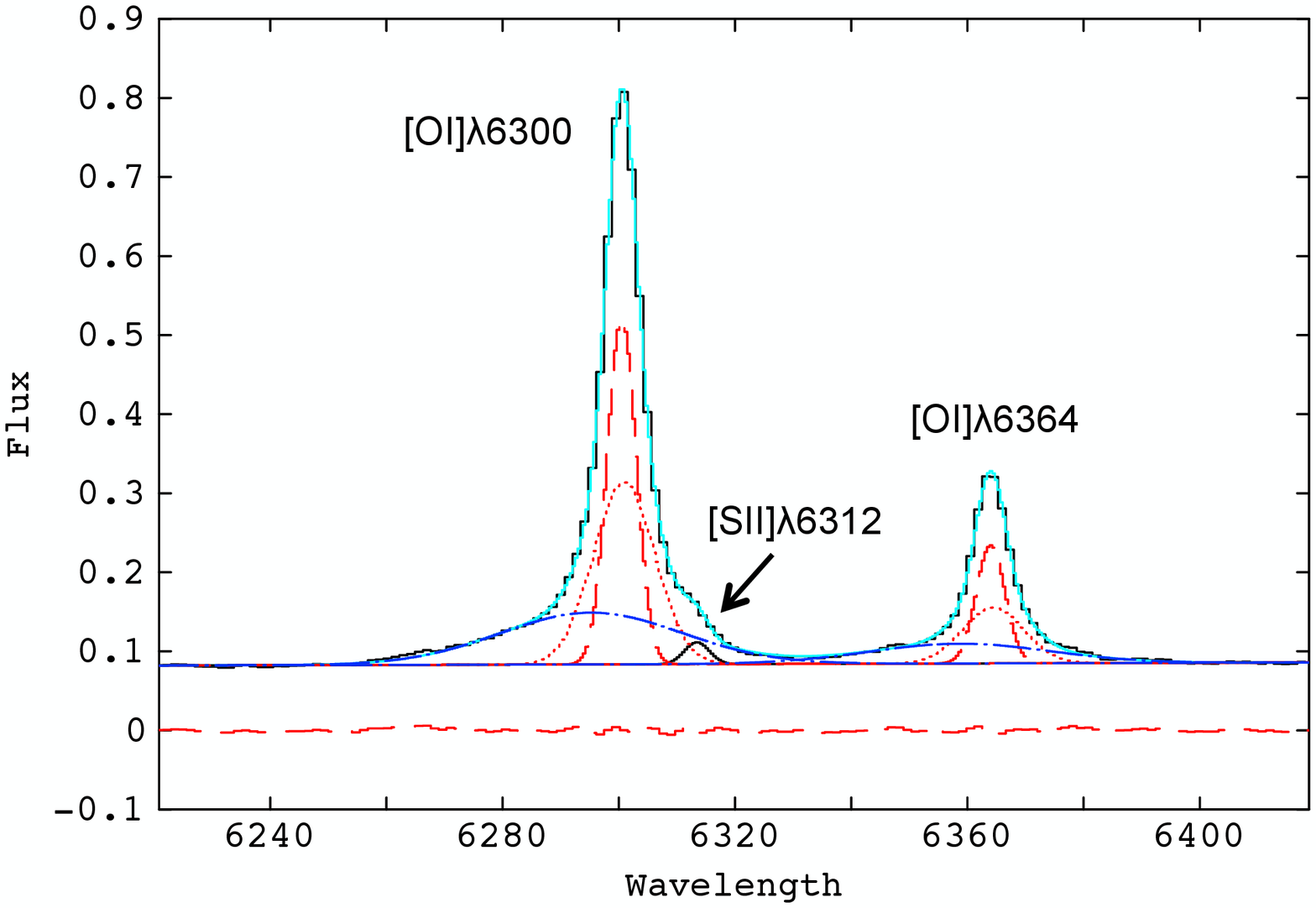}
\vspace{2.5in}
\caption{Spectral fit, residuals  and kinematic components of  [OI]$\lambda\lambda$6300,6364.  [SII]$\lambda$6312 (black small Gaussian) is marginally detected as a
small excess on the red side of the [OI]$\lambda$6300. The detection of [FeX]$\lambda$6375  cannot be confirmed. [OI]$\lambda$6364  is successfully fitted without the contribution of such a line. Color and line style code as in Fig. 5.}
\label{fig:fitoi}
\end{figure}

\begin{figure}
\includegraphics{blue-bump.ps}
\vspace{2.3in}
\includegraphics{red-bump.ps}
\vspace{2.3in}
\includegraphics{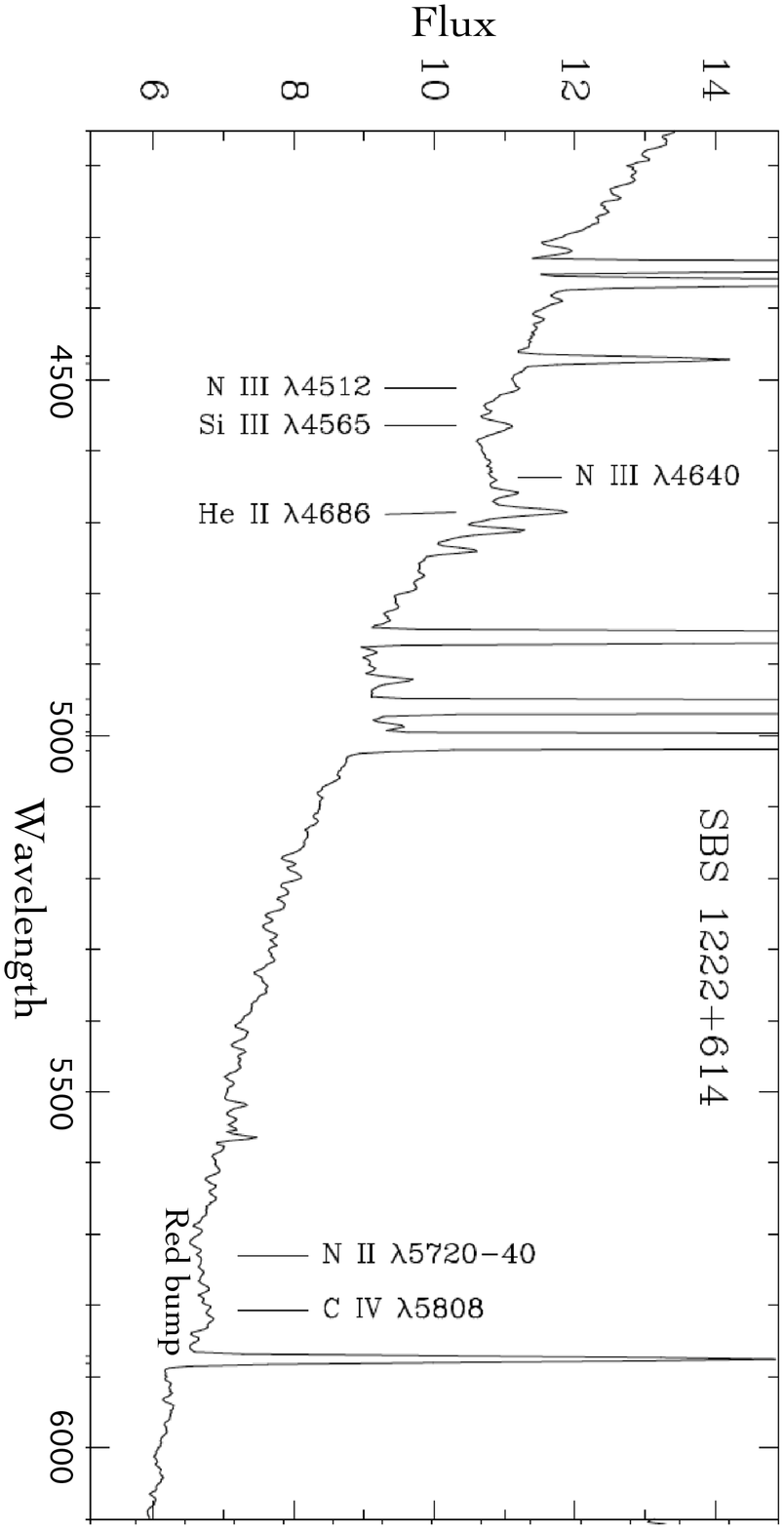}
\vspace{2.1in}
\caption{Top: ``blue bump" near HeII$\lambda$4686 due to WR stars in the SDSS spectrum of MRK~477. Middle: the ``red bump" ($\sim$5800 \AA)  is also  detected.  Bottom: The optical spectrum of the WR galaxy SBS 1222+614  is shown  for comparison (taken from Guseva, Izotov \& Thuan  2000). Flux for this spectrum in units of 10$^{-16}$ erg s$^{-1}$ cm$^{-1}$ \AA$^{-1}$.  }
\label{fig:wr-figure}
\end{figure}

\begin{table*}
\centering
\begin{tabular}{lll|lllllllllll}
\hline
(1) & (2) & (3) & (4) & (5) & (6) & (7) & (8) & (9) & (10)  \\
Comp. & $\frac{\rm H\alpha}{\rm H\beta}$ & $c_{\rm H\alpha}$ & E(B-V)$_{\rm H\alpha}$ & $\frac{\rm H\gamma}{\rm H\beta}$  & $c_{\rm H\gamma}$  & E(B-V)$_{\rm H\gamma}$ &  $\frac{\rm H\delta}{\rm H\beta}$   &  $c_{\rm H\delta}$  &  E(B-V)$_{\rm H\delta}$    \\ \hline
Narrow &  3.1$\pm$0.3   & 0.14$\pm$0.12  & 0.08$\pm$0.08 & 0.47$\pm$0.03 & 0.09$\pm$0.09 & 0.04$\pm$0.03 &  0.23$\pm$0.04 & 0.37$\pm$0.37 &  0.24$\pm$0.11  \\
Intermediate &  4.9$\pm$0.2 &  0.70$\pm$0.04 & 0.47$\pm$0.03 & 0.38$\pm$0.06  & 0.62$\pm$0.46 &  0.42$\pm$0.19 & 0.23$\pm$0.06 & 0.40$\pm$0.40  & 0.26$\pm$0.12    \\
Broad &  4.2$\pm$0.3   & 0.49$\pm$0.08 & 0.32$\pm$0.06 & 0.57$\pm$0.09 & 0.00 &  0.00 & 0.26$\pm$0.03 & 0.15$\pm$0.15 &  0.09$\pm$0.05 \\ \hline
Total &  4.0$\pm$0.2 & 0.45$\pm$0.06 & 0.29$\pm$0.05 & 0.45$\pm$0.05 & 0.21$\pm$0.21 &  0.13$\pm$0.06 & 0.24$\pm$0.02 & 0.27$\pm$0.23  & 0.17$\pm$0.08 \\ \hline
\end{tabular}
\caption{Extinction correction ($c$ and E(B-V)) determined from the Balmer ratios (columns 2, 5 and 8) for the three kinematic components and the integrated lines.}   
\label{tab:reddening}
\centering
\begin{tabular}{lllllllllllll}
\hline
(1) & (2) & (3) & (4) & (5) & (6) & (7)   \\
Comp. & \tiny{$r1^{\rm obs}$=$r1^{\rm int}$} &    $r2^{\rm obs}$  & $r2^{\rm int}$ &  $r3^{\rm obs}$   & $r3^{\rm int}$  & $n$  (cm$^{-3}$) \\ \hline
Narrow &      1.03$\pm$0.03   &   0.05$\pm$0.01  &    0.07$\pm$0.02   &  0.007$\pm$0.005 &  0.010$\pm$0.007 & $\sim$400-630 \\
Intermediate &   0.73$\pm$0.05  & 0.17$\pm$0.02 & 0.37$\pm$0.09  & 0.08$\pm$0.01 & 0.17$\pm$0.04 &  $\sim$2000-4000 \\
Broad &        0.41$\pm$0.12 &  0.24$\pm$0.05 &  0.34$\pm$0.14 & 0.08$\pm$0.07  & 0.25$\pm$0.11 &  $\ga$8000  \\ \hline
Total &   0.76$\pm$0.06 &   0.13$\pm$0.01  &   0.22$\pm$0.03 & 0.05$\pm$0.01 & 0.08$\pm$0.02   & $\sim$1600-3200   \\ \hline
\end{tabular}
\caption{Density  sensitive line ratios observed ($^{\rm obs}$) and corrected ($^{\rm int}$) for reddening. $r1=\frac{\rm [SII]\lambda 6716}{\rm [SII]\lambda 6731}$,
$r2=\frac{\rm [SII]\lambda 4068}{\rm [SII] 6716+6731}$, $r3=\frac{\rm [SII] 4076}{\rm [SII] 6716+6731}$. The range of densities $n$  has been calculated using $r1$. A trend is found  such that the broader the component, the higher the density.}  
\label{tab:temp-dens}
\end{table*}

\vspace{0.2cm}

(ii) {\it Electron density $n$.} We have constrained $n$ with the ratio $r1=\frac{\rm [SII]\lambda 6716}{\rm [SII]\lambda 6731}$ (Osterbrock  \citeyear{ost89}).
The results are shown in Table  \,\ref{tab:temp-dens}.   The  apparently non perturbed ambient gas (the narrow component)  has $n\sim$400-630 cm$^{-3}$. The  intermediate component, which shows intermediate kinematic and physical properties, has $n\sim$2,000-4,000 cm$^{-3}$. Finally, the most extreme kinematic component, which is associated with the outflowing gas
has the highest density $n\ga$8.000 cm$^{-3}$.

We have further checked these results  using   $r2=\frac{\rm [SII] 4068}{\rm [SII] 6716+6731}$ and $r3=\frac{\rm [SII] 4076}{\rm [SII] 6716+6731}$  (Keenan  et al. \citeyear{kee96}). Unlike $r1$, these ratios have a  strong dependence 
on the electron temperature $T_e$ and reddening. They are therefore less efficient at constraining $n$, but
they provide a useful test.
  The dependence  of both ratios with $n$ and  $T_e$
  can be seen in Fig.  \,\ref{fig:keenan}.

To correct for reddening (eq. [2]),  for each component we adopt  $c$ as the average of the maximum and minimum values 
obtained from the three Balmer ratios (Table  \,\ref{tab:reddening}). The uncertainties are calculated using half the difference between these two  values. The results are shown in Table  \,\ref{tab:temp-dens} and Fig.  \,\ref{fig:keenan}.    
    In this figure we show the $n$  sensitive diagnostic diagrams log($r1$) versus  log($r2$)  and  log($r1$) versus  log($r3$) (Keenan  et al. \citeyear{kee96}). The coloured squares mark the areas covered by the allowed range of extinction corrected line ratios   for the different kinematic components and the integrated line fluxes, taking into account all the uncertainties. The figure is consistent with the results
    from $r1$. They support  a trend for increasing density from the narrow to the broad component.

In spite of the uncertainties inherent to fitting the multiple kinematic components in all the lines involved in the determination of $n$  and reddening,
a coherent picture can be drawn: the  ionized gas within a radius of 1.5 arcsec  or $\sim$1.1 kpc (set by the size of the SDSS fibre) 
shows a gradient in physical and kinematic properties, which is apparent in the three kinematic
components. The  apparently non perturbed ambient gas (the narrow component)  has $n\sim$400-630 cm$^{-3}$. The  intermediate component, which shows intermediate kinematic and physical properties, has $n\sim$2000-4000 cm$^{-3}$. Finally, the broadest  component, which is associated with the outflowing gas
has the highest density$\ga$8000 cm$^{-3}$. As mentioned above, the large $\frac{\rm H\alpha}{\rm H\beta}$
decrement might suggest densities as high as $n>$5$\times$10$^{5}$ cm$^{-3}$ for the outflowing gas.

\subsection{Narrow  optical Fe$^+$ emission}

Broad Fe$^+$  multiplet emission are prominent features in the optical spectra of most type 1 AGNs.
	\cite{dong10} have demonstrated statistically that {\it narrow optical} Fe$^+$ lines, either permitted or forbidden, are prevalent in type 1 AGNs, but are completely absent in type 2 AGNs across a wide luminosity range, from Seyfert 2 galaxies to QSO2. 
	
	MRK~477 is an exception.  \cite{hec97}  detected  [FeII]$\lambda$8617 blended with some neighbour lines.    We confirm the detection of  more than 10 Fe$^+$ emission lines (see Table \ref{tab:tablines} and Fig. \ref{fig:iron2}), besides several other features that might be blended or misidentified.   The features for which the width could measured show FWHM$\sim$400-650  km s$^{-1}$, suggesting similar kinematics as the NLR (Fig. \,\ref{fig:kin-corr}, top panels). Other emission lines associated with highly refractory elements are also detected (Mg, Ni and, tentatively, Ca; Figs. \ref{fig:iron2} and  \ref{fig:mag}).

\subsection{Coronal lines}

Coronal lines are  collisionally excited forbidden transitions  emitted within low-lying levels
of highly ionized species  (IP$_{\rm low}>$100 eV; Rodr\'\i guez Ardila et al. \citeyear{rod06}).  According to that exact definition, we cannot confirm the detection of any coronal feature in the {\it optical} spectrum of MRK~477,  since the highest ionization lines we identify unambiguously   are those produced by Fe$^{+6}$  (IP$_{\rm low}=$99 eV).   On the other hand, coronal emission is confirmed in the NIR spectrum
with the detection of  [SiVI]$\lambda$1.963  (IP$_{\rm low}=$167 eV) and   [SiX]$\lambda$1.430 (IP$_{\rm low}=$351 eV) (Fig. \ref{fig:liris}; Table \ref{tab:tabnearir}). 

Although [FeX]$\lambda$6375 (IP$_{\rm low}$=235 eV) was   reported as detected by  \cite{derob87} and  \cite{vei88},  we cannot confirm this.
Both groups used the same spectra in their analysis,  one obtained in 1980 and  another in 1985.  
The 1980 spectrum shows a strong emission feature at the right $\lambda$  (see their Fig. 6, left), but this was discarded by \cite{vei88} as an artifact.
Although the 1985 spectrum shows a small excess  near the [FeX] wavelength  which the authors interpret as the detection
of this line, our analysis shows  that this is consistent with being due to the complex kinematic sub-structure of the [OI]$\lambda$6374 line profile (Fig. \ref{fig:fitoi}).

A broad faint and noisy feature is detected with central $\lambda=$7888.8 \AA. Several lines are possibly
 contributing, including MgII$\lambda$7877, [NiIII]$\lambda$7889; [FeXI]$\lambda$7892 
and MgII$\lambda$7896 (Fig. \ref{fig:mag}, top). Thus, the detection of  [FeXI] emission (IP$_{\rm low}$=262 eV), which has been identified in some type 2 AGN
(Penston et al. \citeyear{pen84}, Rose et al. \citeyear{rose13})  cannot be confirmed either.

The highest  optical ionization lines  unambiguosly identified are [NeV]$\lambda\lambda$3346,3426 (IP$_{\rm low}=$97 eV, out of the SDSS spectral range; reported by De Robertis \citeyear{derob87}) and the [FeVII] lines (IP$_{\rm low}=$99 eV) at $\lambda$ 5159,  6087 \AA~ (also reported by De Robertis \citeyear{derob87}), and lines at 3759, 4893, 5721 \AA. Also  possibly  [FeVII] at 5276 \AA, [ArX] at 5533 \AA, although other identifications cannot be discarded (Table \ref{tab:tablines}).

With somewhat lower ionization level,  [CaV] at 5309 \AA~ (IP$_{\rm low}$=67 eV) and  several [FeVI] (IP$_{\rm low}$=75 eV) lines  at 5146, 5176, 5335, 5485, 5631, 5677 \AA~ are also detected
and  possibly [FeV]$\lambda$5726 (IP$_{\rm low}$=55 eV) and [FeVI] at 5426 \AA.

 [FeVII]$\lambda$6086  is     fainter relative to low ionization lines (e.g.  [OI]$\lambda$6300) than usually found in active galaxies with  strong  coronal emission.     $\frac{\rm [FeVII]\lambda 6086}{\rm [OI]\lambda 6300}$=0.15$\pm$0.01 for MRK~477. For comparison, the Seyfert 1 and 2 galaxies studied by \cite{rod06} with detected [FeX] and [FeXI] lines show  in general (4 out of 5 objects) $\frac{\rm [FeVII]\lambda 6086}{\rm [OI]\lambda 6300}$ in the range $\sim$0.6-5.7.

\subsection{The WR bumps}

\cite{hec97} detected a broad emission complex around He II$\lambda$4686, which is clearly appreciated in the SDSS spectrum.  They fitted 
this so called ``blue bump" satisfactorily as a blend of the He II line together with other lines which are identified in 
Fig. \ref{fig:wr-figure} (top). This unresolved bump is a blend of lines emitted mainly
by late WN and early WC  WR    stars, although some contribution of early WN stars
might be present (Schaerer \& Vacca \citeyear{sch98}). \cite{hec97} 
  propose that it   may be produced by an ensemble of about 30,000 WR stars (WN subtype), in which case  
 MRK~477 would be a luminous (but not extraordinarily so) member of the class 
of WR galaxies.

Another feature often identified in WR galaxies due to WR stars is the so called ``red bump" due to broad CIV$\lambda$5808 emission,
emitted mainly by WCE stars.  This 
is usually much weaker than the ``blue bump" and was not detected by \cite{hec97} . 

 We show in Fig. \ref{fig:wr-figure} (middle panel) the MRK~477 spectrum near the ``red bump"   and 
  the optical spectrum of the WR galaxy SBS 1222+614  (Guseva, Izotov \& Thuan 2000) for comparison (bottom panel).
  The appearance of the ``red bump"  in this and other WR galaxies is very similar to that in  MRK~477.
 We thus believe that the ``red bump"  is detected in the  SDSS spectrum of this QSO2. Its very large width (FWHM $\sim$80 \AA\ or $\sim$4100 km s$^{-1}$)  rules out as its origin the broad wings due to the ionized outflow identified in many other emission lines.

\section{Discussion}

We have analysed the optical  and NIR spectra of MRK~477, the nearest obscured quasar.  WR features first identified by  \cite{hec97} and confirmed here by the new detection of the ``red bump" at $\sim$5800 \AA~ (Sect. 3.7) show  that it has undergone very recent star formation.

The optical+NIR spectrum of MRK~477  is very rich, with $\sim$100 detected emission lines ($\sim$90 in the optical).
In spite of the lack of spatial information of the SDSS spectrum,  the spectral decomposition of numerous lines has allowed us to characterize, at least partially, the spatial structure and the gradients in the physical  and  kinematic properties of the gas. 

 \subsection{The origin  of the FWHM versus  $n_{crit}$ correlation}

As in many other type 1 and type 2 AGN, a significant correlation has been found between the FWHM of the lines and the critical density $n_{\rm crit}$ (in log).
Based on the analysis of  eight type 1 AGN spectra, Stern, Laor \& Baskin (2014; \cite{stern14} hereafter) showed that this relation is consistent with the velocity field within the black hole gravitational sphere of influence  (radius $R_g$), assuming the $n \propto R^{-2}$ relation implied by the equilibrium between the NLR clouds and the radiation pressure  and that the emission of each forbidden line is dominated by gas with $n\sim n_{\rm crit}$.  
This explanation implies that beyond some threshold $n_{\rm crit}$, $n_{\rm crit}^0$, which depends on the luminosity in Eddington units ${\dot m}$, the line emission will be dominated by gas within $R_g$, and therefore is expected to show larger velocities than gas which kinematics is dominated by the host galaxy.   For MRK~477, the stellar velocity dispersion $\sigma_*$=117 \kms (Zhang, Bian \& Wang \citeyear{zhang08}), 
$L_{\rm bol}$ and the values of   $M_{\rm  BH}$ quoted in Sect. 1 imply  $R_g\sim$1  and 40 for log($M_{\rm BH}$)=7.19 and 8.84 respectively (eq. 26 in  \cite{stern14}). This 
implies a  threshold $n_{\rm crit}^0$=10$^{8.6}$ cm$^{-3}$ (eq. 6 in  \cite{stern14}) for log($M_{\rm BH}$)=7.19. This value  is much larger than the $n_{\rm crit}\sim$10$^{4.5}$-10$^{5.5}$ cm$^{-3}$ where the FWHM starts to increase
in MRK 477 (top-left panel in Fig. \ref{fig:kin-corr}), which makes the gravitational interpretation unlikely.  For log($M_{\rm BH}$)=8.84,  $n_{\rm crit}^0$=10$^{5.4}$  cm$^{-3}$  is indeed consistent with the data.  The success of this prediction supports the gravitational interpretation for the FWHM versus $n_{crit}$ correlation, provided MRK~477 harbors such a massive black hole. On the other hand, the kinematic substructure of the lines raises doubts about whether the emission is truly emitted mostly by gas at $n\sim n_{\rm crit}$. As an example, the [OIII] line  flux is dominated by gas with $n\la$4000 cm$^{-3}$ (see Sect. 3.4).

Alternatively,  we propose that the outflow produces this correlation in MRK~477. Thisis supported by  the correlations between $\epsilon = \frac{F_{\rm broad}}{F_{\rm narrow}}$  and $n_{\rm crit}$   (Sect. 3.3.1), which imply that  the outflow  is relatively
stronger  and  is responsible for the increasing line broadening at increasing densities.   This correlation does not differ in any statistical sense from 
the usual FWHM versus  $n_{\rm crit}$ one. The difference lies in the physical interpretation we propose.

It is possible that the effects of ionized outflows explain the FWHM versus  $n_{\rm crit}$ correlation in other AGN.

 Three kinematic components  have been isolated in the main optical  emission lines. The narrowest and intermediate components,  have similar $z$ ($V_S\sim$0 km s$^{-1}$) and are relatively narrow   (FWHM$\sim$[95-210] and [470-560] \kms respectively).
  A third broad blueshifted
 component is moreover isolated in all lines with FWHM in the range $\sim$[1400-1840] km s$^{-1}$ and 
 $V_S\sim$[-490-(-190)] km s$^{-1}$.
 The narrow component is likely to trace the NLR ambient, non perturbed gas, while the broad component is emitted by the most turbulent, outflowing gas. The intermediate component  traces gas of intermediate properties. All three components have line ratios consistent with type 2 active galaxies, as already pointed out by \cite{vm14}, implying that they are spatially located within the quasar ionization cones.

 The difference in kinematics is associated  with a   difference in physical properties. The broader the component, the higher the density. 
 The sequence we find is $n\sim$(400-630) cm$^{-3}$  for the  ambient gas,
 $n\sim$(2000-4000) cm$^{-3}$  for the intermediate component and  $n\ga$8000 cm$^{-3}$  
 for the outflowing gas   (see \S3.3.2). Higher $n$ of the outflowing gas is also suggested by the correlation between  $\epsilon = \frac{F_{\rm broad}}{F_{\rm narrow}}$ and $n_{\rm crit}$  for different forbidden lines (see above).   Density enhancement of the outflowing gas has been found in some radio galaxies and QSO2 (e.g.  Holt et al. \citeyear{holt11}; Villar Mart\'\i n et al. \citeyear{vm14}).

 \subsection{The spatial location of nuclear ionized outflow}

 The isolation of the broad component in  numerous emission lines  demonstrates that the outflow involves gas covering a large range of ionization potentials and critical densities    (at least IP$_{\rm high}\sim$8 to 125 eV, log($n_{\rm crit}) \sim$3.3 to 7.5). We try to constrain  next  the spatial location where it was triggered and how far its effects extend.
 
Bennert et al. (2006a,b) found that the observed gas density of the NLR gas  decreases with  increasing distance to the AGN in Seyfert 1 and 2  galaxies (see also \cite{stern14}).
 A similar behavior  is expected in  MRK~477. The difference in kinematic and physical properties between the broad, intermediate and narrow components 
 can be naturally explained if they are located at increasing distance from the AGN, with the outflow emission dominating at smaller distances, in the inner part of the NLR or even closer (see also Villar-Mart\'\i n et al. \citeyear{vm14}).   The outflow   becomes weaker as it propagates outwards and reaches  less dense regions, where lines of lower critical
densities are preferentially emitted. Because the outflow has lost some power, it drags less and less mass and its emission becomes weaker relative to the ambient gas, which dominates the line fluxes in the most distant, lower density regions. 
This can also explain the correlation of $\epsilon$  with $n_{\rm crit}$.

If the radiation pressure is responsible for the NLR density gradient  (\cite{stern14}),  the estimated $n$  for the three kinematic components   will help to  constrain the distance  from the central engine $R$ at which each one is preferentially emitted.  The expected behaviour of $n$ with the ionizing luminosity  $L_{ion}$
and $R$    is predicted to be (\cite{stern14}): 

$$n = 7 \times 10^4 ~L_{\rm ion,45} ~R_{\rm 50}^{-2}$$

 where $L_{\rm ion,45}$ is  the ionizing luminosity
in units of 10$^{45}$ erg s$^{-1}$ and $R_{\rm 50}$ is the distance in units of 50 pc.

Typical  $\frac{L_{\rm bol}}{L_{\rm ion}}$ ratios  for luminous Sy1 and QSO1 with $L_{\rm bol}$ in the range $\sim$10$^{45-46}$ erg s$^{-1}$ have  median  value of 3.89 and standard deviation 4.36. Thus, for a ratio of 3.89, then   $L_{\rm ion,45}\sim$2.2 for MRK~477. Considering the range of $n$ inferred for each gaseous component, the broad (outflowing), intermediate and narrow  kinematic components would be located at  $\la$220 (the upper limit is a consequence of the lower limit on $n\ga$8000 cm$^{-3}$), 375$\pm$65 and 880$\pm$100 pc  respectively.  
This sets an upper limit on the distance at which the outflow has been originated of $R\la$220 pc. If densities as high as $\sim$5$\times$10$^5$
cm$^{-3}$   exist in the outflowing gas (Sect. 3.4.2), then $R\la$30 pc. For comparison, the radio jet extends at $\sim$0.6 arcsec  or 435 pc from the central engine. With all  uncertainties involved, it is remarkable   that these calculations place  the intermediate and narrow components  near and  beyond the edge of the radio  jet respectively.  This adds  further support to the idea that the radio jet has originated the outflow. In this scenario, our results suggest that  the outflow in MRK~477 is concentrated in the nuclear region and does not reach distances beyond $\sim$few$\times$100 pc.

 Alternatively, the higher density of the outflowing gas might be explained by the compression exerted on the outflowing gas by the radio-jet induced shocks (Villar Mart\'\i n et al. \citeyear{vm99}; Holt et al. \citeyear{holt11}). We note that this  scenario is qualitatively different from the scenario where the NLR clouds are compressed by radiation pressure, as assumed in the models used above.  The higher density     would 
 instead be located at or near the radio jet and decrease moving away from it.  
  In this case a correlation  would be expected between the morphology of the radio jet and the ionized gas (whose emission is enhanced due to  shock excitation and/or  density enhancement). This is actually the case in MRK~477  (Heckman et al. \citeyear{hec97}).  Both the radio and [O III] images
show a bright knot or ridge of emission about 0.4 arcsec ($\sim$290 pc) to the northeast of the central source. This clearly shows that the radio jet is interacting with the NLR and enhancing the emission at this location.  Thus, an alternate scenario is that this detached knot is responsible for the bulk of emission of the outflowing gas. If so,  it must emit a wide range of  lines, including [FeVII]$\lambda$6087.
High spatial resolution spatially extended optical spectroscopy would help discriminate between both scenarios.

\subsection{The origin of the NIR narrow [FeII]$\lambda$1.644 emission}

 We have found that this line  is  much broader  than expected for its critical density and ionization potentials (Sect. 3.2). The  underlying
broad component has  FWHM$\sim$4770$\pm$830 km s$^{-1}$, larger than any other line  and contributes more than half of the total line flux ($\sim$57\%). It  also shows the largest $\epsilon=$1.3$\pm$0.2 and
is the only line for which the broad component is not blueshifted. All these could be naturally explained in terms of reddening being much weaker in the NIR, so that both the approaching and receding parts of the expanding outflow are observed (unlike the optical lines). However, this is an unlikely  explanation since Pa$\alpha$ shows that the broad component  is  blueshifted  by -570$\pm$119 km s$^{-1}$.

 Alternatively, it is possible that  [FeII]$\lambda$1.644 is relatively enhanced  by the shocks induced by the outflow,   more than other lines and at a less obscured spatial location (hence the non-blueshift).   Outflow induced shocks  have been often proposed as a relevant [FeII]$\lambda$1.644 excitation   mechanism  in active galaxies (e.g. Ramos Almeida et al. \citeyear{ram09}, Contini et al. \citeyear{con12}).  As an example, \cite{rod04} studied the kinematics and excitation mechanism of the NIR emission lines in a sample of Seyfert 1 galaxies, including  [FeII]
$\lambda$1.257 and $\lambda$1.645 $\mu$m. They found 3  out of 22 sources  in which the [FeII] lines were the broadest, even broader than the coronal  [SiX]$\lambda$1.252 $\mu$m. They suggest that in these objects [FeII] must arise from and additional source, partially formed in a region distinct
from other low-ionization species, which they suggest to be 
associated with  shock excitation from the radio jet.   \cite{bli94} also suggest that the  [FeII]$\lambda$1.644  emission in the Sy2 galaxy NGC1068 is preferentially emitted by gas which is interacting with the nuclear outflow/jet  and is possibly located at the interface between the  outflow and the
dense circumnuclear molecular clouds.

 \subsection{Coronal lines}

 Detection of coronal emission is confirmed in the NIR spectrum, but not in the optical  (Sect. 3.6). In the optical, 
the highest ionization lines are those emitted by Fe$^{+6}$, which on the other hand are  fainter relative to the low ionization lines than typically found in AGN with  strong coronal emission. Coronal lines are generally detected between just a few parsecs  and a few hundred parsecs (e.g. Mazzalay et al. \citeyear{maz10};\citeyear{maz14}). They  have been proposed to be formed in a region located at an intermediate distance between the classical NLR and the broad-line region (BLR) (M\"uller S\'anchez et al. \citeyear{mul06}).  Alternatively, the coronal region might reside in the inner wall of the dusty torus (Murayama \& Taniguchi \citeyear{mur98}, Rose et al. \citeyear{rose13}) or in the low density surface layer of the radiation pressure confined NLR clouds (\cite{stern14}).

The detection of coronal emission in the NIR and its absence in  the optical spectrum of MRK~477  suggests that this region is heavily reddened, partially    hidden from our line of sight.   Alternatively, since the coronal lines are usually highly nucleated ($<$1 kpc from the nucleus) and much more concentrated than lower ionization NLR  features (Rodr\'\i guez Ardila et al. \citeyear{rod11}), it is also possible that they are diluted by the strong continuum contribution within the large  aperture of the SDSS fibre (radius $\sim$1 kpc). 

The [SiVI] and [SiX] lines have luminosities 40.0 and 39.8 in log and erg s$^{-1}$respectively.  \cite{rod11}
found that these lines display a narrow range in luminosity in Seyfert 1 and 2, with values for most objects located 
in the interval log($L$)=39-40.  MRK~477 is at the high end of this range. We measure [SiVI]/[SiX]=1.6$\pm$0.1, also within the range  measured for Seyfert 1 and 2 galaxies by those authors.

\subsection{Detection of narrow optical Fe$^+$ emission}

We have identified more than 10  {\it narrow optical} Fe$^+$ emission lines  in the SDSS spectrum of MRK~477.   The NIR 
[FeII] lines at 1.26 and 1.64 $\mu$m are routinely detected in type 2  Seyferts (e.g. 	Ramos Almeida, P\'erez Garc\'\i a \& Acosta-Pulido 
\citeyear{ram09}). However,  to our knowledge MRK~477 is the first type 2 AGN with optical Fe$^+$ line  detections. 

To explain the  absence of narrow optical Fe$^+$  in type 2  AGN, while being prevalent in type 1 objects, \cite{dong10} 
proposed that such   emission  is confined to a disk-like geometry in the innermost region of the NLR on physical scales of parsecs.
This would be smaller than the obscuring torus and within the dust sublimation radius.   Iron, which is a refractory element which easily condenses on to dust grains, is in the gaseous  phase in the absence of dust (Laor \& Draine \citeyear{laor93}). 
In this scenario  the narrow  optical Fe$^+$ emitting region is visible along our line of sight in type 1 objects, but obscured by the extent of the dusty torus in type 2 counterparts.  

It is  however difficult to picture a geometry such as this to explain the optical Fe$^+$  emission in MRK~477. Broad permitted   hydrogen and helium lines
 would be expected, unless a BLR did not exist, but this is not the case  (Sect. 1).

We explore next whether the optical Fe$^+$ features can be naturally explained by the intrinsic NLR emission. For this, we compare the measured integrated flux relative to H$\beta$ 
of all Fe$^+$ lines in the blue (4000-6000 \AA) and red (6000-7800 \AA) bands with the ratios predicted by photoionization models appropriate for the NLR conditions. 
We use the radiation-pressure-confined NLR models described in \cite{stern14}. The assumption of these models  implies that the line emission is essentially independent of the ionization parameter at the slab surface $U_0$, as long as $U_0\gg0.03$ (see Fig.~3 in S14; see also Dopita et al.  \citeyear{dop02} and  Groves et al. \citeyear{gro04}). A continuous distribution of dusty gas as a function of the distance from the nucleus $R$ is considered. This distribution is characterized by $\eta$, the power-law index of the dusty gas covering factor as a function of logarithmic unit of $R$. These types of models are successful in explaining various observations of the NLR (see Dopita et al.\ 2002 and S14). 

We  assume solar metallicity, an ionization slope of $\alpha_{\rm ion}=-1.6$ typical of luminous quasars (Telfer et al. \citeyear{tel02}), and $\eta=0$, i.e.\ a constant covering factor per $\log\ R$, as suggested by the strong line ratios and the IR spectral energy distribution (S14). The photoionization models are calculated using {\sc cloudy} (Ferland et al.~\citeyear{fer13}), assuming hydrostatic equilibrium.  Depletion of refractory elements on to dust grains is taken into account, and the Fe$^+$ ion is calculated according to the model described in Verner et al.\ (1999).

The results for the ${\rm FeII}(4000-6000)/{\rm H}\beta)$ and ${\rm FeII}(6000-7800)/{\rm H}\beta)$ are shown in Table \ref{tab:femodels} considering both dusty and non-dusty models. The uncertainties on the observed ratios account for the errors on the  flux measurements and the possible misidentification of some features (Table \ref{tab:tablines}). The models can reproduce successfully both the blue  and the red bands. Therefore, the narrow optical Fe$^+$ lines in MRK~477 can be naturally explained by the intrinsic NLR emission due to AGN photoionization. The high luminosity and/or closeness of MRK~477  could be the reason why these lines have been detected, unlike other  type 2 AGN.

The effect of dust depletion produces ratios approximately twice lower than those derived from dust-free  models.
 Taking into account the uncertainties, the Fe$^+$ band ratios do not discriminate between the two scenarios. Model predictions of lines emitted by other refractory elements such as Mg, Ni, Ca (also detected for MRK~477, see Sect. 3.5)  would help to infer whether the NLR contains dust-free gas.  
Only the [CaII]$\lambda$7291 is predicted by {\sc cloudy}. Calcium is much more sensitive to dust depletion than iron. 
The dust-free models predict [Ca II]$\lambda$7291/H$\beta$=0.08, while the line would be 1000 times fainter (and thus completely undetectable) in the dusty case (see also Villar Mart\'\i n et al. \citeyear{vm96}). The [Ca II] line is tentatively detected with a ratio of 0.02$\pm$0.01 relative to H$\beta$ (Fig. \ref{fig:iron2}, third panel). If this detection is confirmed, it would imply that at least a fraction of the NLR gas is dust free. 

\begin{table*}
\centering
\begin{tabular}{cccccccc}
\hline
(1) & (2) & (3)  & (4) & (5) & (6) \\ 
$\big (\frac{\rm FeII_{\rm 4000-6000}}{\rm H\beta}\big)^{\rm obs}$ 	&  $\big (\frac{\rm FeII_{\rm 4000-6000}}{\rm H\beta}\big)^{\rm int}$	&    Model$_1$  & Model$_2$ &
$\big (\frac{\rm FeII_{\rm 6000-7800}}{\rm H\beta}\big)^{\rm obs}$  	&  $\big (\frac{\rm FeII_{\rm 6000-7800}}{\rm H\beta}\big)^{\rm int}$ &  Model$_1$  & Model$_2$ \\   
& &   Dust & Dust-free  & &    & Dust & Dust-free   \\ \hline  
 0.14$\pm$0.03 & 0.15$\pm$0.1    & 0.15 & 0.36 & 0.06$\pm$0.03 & 0.05$\pm$0.04   & 0.031 & 0.053  \\
\hline 
\end{tabular}
\caption{Comparison between the measured integrated flux relative to H$\beta$ 
of all Fe$^+$ lines in the blue (4000-6000 \AA) and red (6000-7800 \AA) bands with the ratios predicted by photoionization models appropriate for the NLR conditions (see the text). The superscripts $^{\rm obs}$ and $^{\rm int}$  refer to the observed and reddening corrected ratios. Dusty and dust-free model predictions are shown. The narrow optical Fe$^+$ lines in MRK~477 can be naturally explained by the intrinsic NLR emission due to AGN photoionization.}  
\label{tab:femodels}
\end{table*}

\section{Conclusions}

We perform a detailed spectral analysis and characterization of  the NLR   of the nearest obscured quasar (QSO2) MRK~477  
at $z$=0.037  based on the  SDSS optical and NIR H+K spectra obtained with WHT-LIRIS. 

(i)  We confirm the hybrid nature (AGN+starburst) proposed by  Heckman et al. \citeyear{hec97}, based on the new detection of  the  ``red bump" at $\sim$5800 \AA ~due to WR stars, which implies that the system has undergone recent star formation.

(ii)  The optical spectrum of MRK~477  is rich in emission lines, with $\sim$90 detected features. 
   In spite of the lack of spatial information of the SDSS spectrum, the spectral decomposition of numerous lines has allowed us to characterize, at least partially, the spatial structure and the gradients in the physical and kinematic properties of the gas.

(iii) Gas densities within the range $\sim$1945$^{+670}_{-460}$ and up to $>$10$^4$ cm$^{-3}$ are confirmed in the NLR of MRK 477.

(iv)  As in many other active galaxies (AGN), a significant correlation is found between the lines FWHM and the critical density log($n_{\rm crit}$).
 We propose that this correlation is caused by the outflow and its impact on the gas kinematics. This could be the case in other AGNs.

(v)   MRK~477 is an example of a  radio quiet  powerful AGN where negative feedback (the nuclear outflow)  can be dominated by the radio structures.
The  outflow emission has been isolated in many emission lines  covering a large range of ionization potentials and critical densities (from [OI]$\lambda$6300 to [FeVII]$\lambda$6086). The outflowing gas, which is concentrated within R$\sim$few$\times$100 pc from the central engine,  is $>$13 times denser ($n\ga$8,000 cm$^{-3}$) than the ambient non perturbed gas ($n\sim$400-630 cm$^{-3}$). Marginal evidence is found for densities as high as $n\ga$5$\times$10$^5$ cm$^{-3}$ in the outflowing gas. It is possible that the density enhancement is due to the gas compression produced by   jet induced shocks. This is supported by the correlation between the radio and [OIII]$\lambda$5007 morphologies found by \cite{hec97}.

Alternatively, the density enhancement is not related to the jet. Instead, it might be  a reflection   of  the NLR intrinsic density gradient, consequence of the gas being compressed by radiation pressure. In this scenario, and  based on the comparative study between the density and the kinematic properties of the outflowing and the ambient gas,  we conclude that the outflow has been generated  at $\la$220 pc (possibly at $\la$30 pc) from the AGN. We find evidence of  how its effects weaken  as it    propagates outwards, following the NLR density gradient. Beyond the radio jet edge,  the gas emission is dominated by ambient less dense, non perturbed gas. This adds further support to the idea that the radio jet has triggered the outflow.

(vi)  The [FeII]$\lambda$1.644$\mu$m line presents a very different behaviour than the rest of the emission lines. It shows the most extreme effects of the outflow, with an underlying broad component of FWHM=4770$\pm$830 km s$^{-1}$. Its properties suggest that its emission is enhanced by shocks induced by the nuclear outflow/jet and is preferentially emitted  at a different, less reddened spatial location, maybe the interface between the  outflow and the dense circumnuclear molecular clouds.

(vii)  More than 10 {\it narrow optical} Fe$^+$ emission lines have been detected in the SDSS spectrum of MRK~477. To our knowledge, this is the first type 2 AGN with such a detection. We show that these lines can be explained as the natural emission from  NLR gas photoionized by the AGN. Emission lines associated with other highly refractory elements (Mg, possibly Ni) are also detected.  If the tentative detection of the [Ca II]$\lambda$7291 line is confirmed, this would imply that at least part of the NLR gas is dust-free. 

(viii) Coronal line emission is confirmed in the NIR, but not in the optical SDSS spectrum.
The coronal region might be heavily reddened, partially hidden from our line of sight. Alternatively its optical emission  might be    diluted  due to  the large SDSS fibre aperture. The coronal region also participates in the outflow.
 
(ix)  Pa$\alpha$ is spatially extended along the radio and [OIII] emission axes, up to a maximum radial extension of $\sim$1.5 kpc from the AGN.
  
\section*{Acknowledgments}

Thanks to an anonymous referee for useful comments on the paper.

This work has been funded with support  from the Spanish Ministerio de Econom\'\i a y Competitividad through  the grant AYA2012-32295. JS acknowledges financial support from the Alexander von Humboldt foundation. CRA is supported by a Marie Curie Intra European Fellowship within the 7th European Community Framework Programme (PIEF-GA-2012-327934). RGD acknowledges support through the grant AYA2010-15081.

This work  is partially based   on data obtained as part of the WHT service programme. The  WHT and its service programme are operated on the island of La Palma by the Isaac Newton Group in the Spanish Observatorio del Roque de los Muchachos of the Instituto de Astrof\'\i sica de Canarias. We thank the WHT staff for performing the observations.

The work is also based on data from Sloan Digital Sky Survey. Funding for the SDSS and SDSS-II has been provided by the Alfred P. Sloan Foundation, the Participating Institutions, the National Science Foundation, the U.S. Department of Energy, the National Aeronautics and Space Administration, the Japanese Monbukagakusho, the Max Planck Society, and the Higher Education Funding Council for England. The SDSS website is http://www.sdss.org/. 

The SDSS is managed by the Astrophysical Research Consortium for the Participating Institutions. The Participating Institutions are the American Museum of Natural History, Astrophysical Institute Potsdam, University of Basel, University of Cambridge, Case Western Reserve University, University of Chicago, Drexel University, Fermilab, the Institute for Advanced Study, the Japan Participation Group, Johns Hopkins University, the Joint Institute for Nuclear Astrophysics, the Kavli Institute for Particle Astrophysics and Cosmology, the Korean Scientist Group, the Chinese Academy of Sciences (LAMOST), Los Alamos National Laboratory, the Max-Planck-Institute for Astronomy (MPIA), the Max-Planck-Institute for Astrophysics (MPA), New Mexico State University, Ohio State University, University of Pittsburgh, University of Portsmouth, Princeton University, the United States Naval Observatory, and the University of Washington.

\end{document}